\documentclass[twocolumn]{aastex701} 

\usepackage{multirow}
\usepackage{amsmath}
\usepackage{apjfonts} 
\usepackage{comment}
\usepackage{enumitem}
\usepackage{CJK}

\newcommand{\mmicron}{$\mu$m}
\newcommand{\spitzer}{\textit{Spitzer}}
\newcommand{\herschel}{\textit{Herschel}}
\newcommand{\dynesty}{\textsc{dynesty}}
\newcommand{\prospector}{\textsc{prospector}}
\newcommand{\cigale}{\textsc{cigale}}

\newcommand{\jwst}{\textit{JWST}}

\newcommand{\hb}{\hbox{H$\beta$}}
\newcommand{\ha}{\hbox{H$\alpha$}}

\newcommand{\oiii}{[\ion{O}{3}]}

\submitjournal{ApJ}

\shorttitle{Spectrophotemetric SED Fitting of Little Red Dots}
\shortauthors{Ronayne et al}

\begin{document}
\begin{CJK}{UTF8}{gbsn}
\title{MEGA: Spectrophotometric SED Fitting of Little Red Dots Detected in JWST MIRI}

\author[0000-0001-5749-5452]{Kaila Ronayne}
\affiliation{Department of Physics and Astronomy, Texas A\&M University, College Station, TX, 77843-4242 USA}
\affiliation{George P.\ and Cynthia Woods Mitchell Institute for Fundamental Physics and Astronomy, Texas A\&M University, College Station, TX, 77843-4242 USA}
\email[show]{kaila\_ronayne@tamu.edu}

\author[0000-0001-7503-8482]{Casey Papovich}
\affiliation{Department of Physics and Astronomy, Texas A\&M University, College Station, TX, 77843-4242 USA}
\affiliation{George P.\ and Cynthia Woods Mitchell Institute for Fundamental Physics and Astronomy, Texas A\&M University, College Station, TX, 77843-4242 USA}
\email{papovich@tamu.edu}

\author[0000-0002-5537-8110]{Allison Kirkpatrick}
\affiliation{Department of Physics and Astronomy, University of Kansas, Lawrence, KS 66045, USA}
\email{akirkpatrick@ku.edu}

\author[0000-0001-8534-7502]{Bren E. Backhaus} 
\affil{Department of Physics and Astronomy, University of Kansas, Lawrence, KS 66045, USA} 
\email{bren.backhaus@ku.edu}

\author[0000-0002-3736-476X]{Fergus Cullen}
\email{fergus.cullen@ed.ac.uk}
\affiliation{Institute for Astronomy, University of Edinburgh, Royal Observatory, Edinburgh EH9 3HJ, UK}
\email{fergus.cullen@ed.ac.uk}

\author[0000-0001-9495-7759]{Lu Shen}
\affiliation{Department of Physics and Astronomy, Texas A\&M University, College Station, TX, 77843-4242 USA}
\affiliation{George P.\ and Cynthia Woods Mitchell Institute for Fundamental Physics and Astronomy, Texas A\&M University, College Station, TX, 77843-4242 USA}
\email{lushen@tamu.edu}

\author[0000-0002-9921-9218]{Micaela B. Bagley}
\affiliation{Department of Astronomy, The University of Texas at Austin, Austin, TX, USA}
\affiliation{Astrophysics Science Division, NASA Goddard Space Flight Center, 8800 Greenbelt Rd, Greenbelt, MD 20771, USA}
\email{mbagley@utexas.edu}

\author[0000-0001-6813-875X]{Guillermo Barro}
\affiliation{University of the Pacific, Stockton, CA 90340 USA}
\email{guillermobc@gmail.com}

\author[0000-0001-8519-1130]{Steven L. Finkelstein}
\affiliation{Department of Astronomy, The University of Texas at Austin, Austin, TX, USA}
\email{stevenf@astro.as.utexas.edu}

\author[0000-0002-6292-4589]{Kurt Hamblin}
\affiliation{Department of Physics and Astronomy, University of Kansas, Lawrence, KS 66045, USA}
\email{kurt.hamblin@ku.edu}

\author[0000-0001-9187-3605]{Jeyhan S. Kartaltepe}
\affiliation{Laboratory for Multiwavelength Astrophysics, School of Physics and Astronomy, Rochester Institute of Technology, 84 Lomb Memorial Drive, Rochester, NY 14623, USA}
\email{jsksps@rit.edu}

 \author[0000-0002-8360-3880]{Dale D. Kocevski}
\affiliation{Department of Physics and Astronomy, Colby College, Waterville, ME 04901, USA}
\email{dkocevsk@colby.edu}

\author[0000-0002-6610-2048]{Anton M. Koekemoer} 
\affiliation{Space Telescope Science Institute, 3700 San Martin Drive, Baltimore, MD 21218, USA} 
\email{koekemoer@stsci.edu}

\author[0000-0003-3216-7190]{Erini Lambrides}\altaffiliation{NPP Fellow}
\affiliation{NASA-Goddard Space Flight Center, Code 662, Greenbelt, MD, 20771, USA}
\email{elambrid@gmail.com}

\author[orcid=0000-0001-9879-7780]{Fabio Pacucci}
\affiliation{Center for Astrophysics $\vert$ Harvard \& Smithsonian, 60 Garden St, Cambridge, MA 02138, USA}
\affiliation{Black Hole Initiative, Harvard University, 20 Garden St, Cambridge, MA 02138, USA}
\email{fabio.pacucci@cfa.harvard.edu}

\author[0000-0001-8835-7722]{Guang Yang(杨光)}\altaffiliation{Current Institution: Nanjing Institute of Astronomical Optics and Technology, Nanjing 210042, China }
\affiliation{Kapteyn Astronomical Institute, University of Groningen, P.O. Box 800, 9700 AV Groningen, The Netherlands}
\affiliation{SRON Netherlands Institute for Space Research, Postbus 800, 9700 AV Groningen, The Netherlands}
\email{gyang206265@gmail.com}

\begin{abstract} 
We analyze eight spectroscopically confirmed Little Red Dots (LRDs) at redshifts $z = 5.1-8.7$ with \jwst/NIRCam, NIRSpec, and MIRI data. The LRDs have red NIRCam colors, F150W--F444W $>$ 1, but flat NIRCam--MIRI colors, $-0.5 < \mathrm{F444W - F770W} < 0.5$, suggesting weak warm/hot dust components. The LRDs have $-1.0 < \mathrm{F1000W - F1500W} < 1.1$, suggestive of non-uniform rest near-IR properties within the sample. We model the spectral energy distributions (SEDs) of the LRDs using the \textsc{CIGALE} and \textsc{Prospector} codes to assess how the differing templates impact the interpretation for LRDs for cases of: (1) models with star-forming stellar populations only; (2) active galactic nuclei (AGN) dominated models; and (3) composite AGN and star-forming models. Using the Bayesian information criterion, we find that six of the eight LRDs favor AGN models compared to star-forming models, though no model reproduces all of the observed properties. Two LRDs with pronounced Balmer-breaks and broad \ha\ have SEDs that are reproduced with hot, dense-gas ($\log T/\mathrm{K}=5-5.7$, $\log n/\mathrm{cm^{-3}} = 9-11$) models with low dust attenuation ($A(V)\simeq 0.5$~mag). However, these models require an additional thermal component (800-1400~K) to account for the MIRI data, and fail to reproduce the rest-UV and narrow \oiii\ emission.  The total bolometric emission from the dense-gas models, and possibly CIGALE AGN models, appear consistent with literature constraints in the far-IR and radio, and require $\log L_\mathrm{bol}/L_\odot<12$. These results suggest that our LRDs cannot be modeled entirely with standard templates, but instead require a novel treatment of gas conditions, AGN and star-formation.
\end{abstract}


\section{Introduction} \label{sec:intro}
Since their discovery with \jwst, Little Red Dots (LRDs) have been the subject of extensive discussion in the literature due to their unprecedented properties, with no clear analog in the nearby universe. LRDs manifest as point-like sources in the reddest NIRCam bands, show red NIRCam colors, blue rest-UV continua, and predominantly reside at $z{>}5$ \cite{kocevski2024,Kokorev_2024,Labbe2025}. Spectroscopic observations of these objects show that some LRDs have broad Balmer emission lines, with FWHM as high as 5000~km~s$^{-1}$ \cite{Greene_2024, Matthee_2024}. These properties are indicative of an AGN  \citep{Harikane_2023, Kocevski2023, wang2024}, as the observed line widths are significantly broader than those typically found in local star-forming galaxies \citep{Fumagalli2012}, and the presence of narrow forbidden lines (e.g., [\ion{O}{3}]) argues against outflows as the dominant source of the broadening. However, the weak/absent X-ray emission \citep{Ananna_2024, Yue_2024} and missing high ionization lines \citep{Akins_2024,Lambrides_2024} suggests that their nature might be more complex. \par
 
Observations from \jwst/MIRI provide a deeper understanding of LRD properties, revealing insights on their dust that cannot be determined with NIRCam alone. MIRI constraints in spectral energy distribution (SED) models for large samples generally favor a compact, dust-rich starburst with minimal AGN contribution, where the averaged SED beyond 5 \mmicron\ was consistent with the expected emission from star forming stellar populations \citep{pablo2024}. In contrast, the stacked SEDs of LRD candidates with MIRI data from \cite{Williams2024} showcases a clear absence of hot dust, indicating the presence of older stellar populations with a potential contribution from an obscured AGN.  \cite{wang2024} reported a broad-line LRD lacking hot dust at $z=3.1$ with MIRI detections, proposing that its UV/optical emission is dominated by an AGN without a torus, underscoring the unsettled nature of LRDs. Indeed, the most favorable solution seems to be some combination of stellar light and AGN emission when the MIRI data is included in the fits \citep{Barro2024, Leung2024}.\par 

While the SED results appear to represent the emissions of LRDs, persistent uncertainties in the models make it difficult to fully discern their physical nature. Current star-forming-only models or AGN-only models struggle to describe the characteristic ''v'' shaped SED \citep{labbe2023}, which can influence the model preference towards a composite solution \citep{Furtak_2023,Vidal2024,Williams2024,Ma2025}. Alternative emission processes for LRDs have been proposed, such as compact low-metallicity starbursts \cite{Baggen_2024, Kokubo2024} with patchy dust \citep{Akins2023, Kocevski2023, Barro2024, Killi2024, Labbe2025}, AGN with super-Eddington accretion \citep{Inayoshi_2024,Lambrides_2024, Madau_2024,Pacucci_2024}, inelastic Raman scattering \citep{Kokubo_2024}, or black holes surrounded by dense, warm-gaseous outflows \citep{degraaff2025,Naidu_2025, taylor2025}, none of which have been ruled out yet.\par

Addressing the nature of LRDs requires a multi-pronged approach, entailing both the study of spectra and photometric data from a diverse sample spanning as much of the SED as possible, and a comprehensive analysis of the SED models.  The inclusion of the \jwst/MIRI data, which are sensitive enough to detect LRDs at 7.7-21 \mmicron, helps to distinguish the stellar from the AGN emission \citep{Kirkpatrick2023,Casey_2024,Yang2023}. At these wavelengths we expect the differences in the (rest-frame) near-IR continuum and emission features from star-formation and AGN to be present.\par

In this paper we present results for a sample of eight LRDs using \jwst/MIRI, NIRSpec/prism, and NIRCam data to interpret spectral energy distribution (SED) models with \cigale\ \citep{Boquien_2019} and \prospector\ \citep{Leja_2019,Johnson2021}, allowing us to test systematics between the results. To do this we use multi-wavelength photometric catalogs from the Cosmic Evolution Early Release Science (CEERS) Survey \citep[][]{finkelstein2017}, combined with MIRI imaging from the MIRI EGS Galaxy and AGN Survey \citep[MEGA][]{backhaus2025}, and NIRSpec prism data from Red Unknowns: Bright Infrared Extragalactic Survey \citep[RUBIES,][]{deGraaff2024} and CEERS \citep{ArrabalHaro2023}. We explore a range of models, including pure star-forming stellar populations, AGN dominated models, and AGN and star-forming composite models. The \cigale\ and \prospector\ templates differ in their treatment of the components in each model, enabling us to assess the systematic uncertainties introduced by these modeling choices.  We also consider dense-gas models for sources with strong Balmer breaks. We then compare the resulting properties across the various LRDs in our sample.

The outline of this paper is as follows. Section \ref{sec:data} 
provides an overview of our \jwst/MIRI, NIRCAM, \textit{Spitzer}/MIPS, and Hershcel/PACS imaging and catalogs as well as our NIRSpec G395M and prism data. We describe our selection methods and sample in Section \ref{sec:sample}. Section \ref{sec:methods} describes our SED modeling methods and measurement of the BIC used to determine the best-fit solutions. In Section \ref{sec:results} we show our results and then discuss their implications and caveats in Section \ref{sec:discussion}, where we also discuss alternative models, such as the dense-gas models with additional thermal components. Lastly, our summary and main conclusions are in Section \ref{sec:summary_concl}. We use the standard Lambda cold dark matter (\(\Lambda\)CDM) cosmology with \(H_0\) = 70 km \({Mpc}^{-1} s^{-1}\), \(\Omega_\Lambda\) = 0.70, and \(\Omega_M\) = 0.30. Throughout this paper all magnitudes are presented in the AB system \citep[][]{Oke1983, Fukugita1996}.

\section{Data} \label{sec:data}
\subsection{MIRI Imaging and Photometric Catalog}
For our MIRI data in this study we use the MEGA survey (PID: 3794; PI: A. Kirkpatrick) which overlaps with the CEERS survey \citep[][PID: 1345]{finkelstein2017} to add MIRI coverage to the Extended Groth Strip \citep[EGS,][]{davis2007} with the F770W, F1000W, F1500W, and F21000W MIRI filters. A full description of the MIRI data reduction as well as the MEGA survey configuration can be found in \cite{backhaus2025}, but we provide some details below. 

The MIRI images were processed with the \textsc{JWST Calibration Pipeline} \citep[v1.14.0;][]{Bushouse2024} with additional steps to mask warm pixels and perform custom background subtraction. The default parameters are used for stages 1 and 2; however the background subtraction step was skipped. Median stacking was then done for the resulting ``.cal'' files per filter from stage 2, which followed the procedure from \cite{alberts2024} to mask warm pixels. Custom background subtraction was then implemented once again following the procedure from \cite{alberts2024}, where the warm pixel corrected ``cal'' files were pushed through stage 3 of the pipeline without the background or \textit{tweakreg} steps. A source mask is then generated from the background subtracted images, which are then mapped back onto the warm pixel corrected individual cal.fits files. These are then corrected for the vertical/horizontal striping using the method in \cite{yang2023b}, which are put back into the stage 3 pipeline two more times using the same settings. The final mask of all sources is mapped back onto the original un-filtered warm pixel corrected cal.fits files, where the median vales for the masked warm pixels are subtracted to bring the median background value to zero. With the un-filtered warm pixel corrected cal.fits file, a super background is created by stacking the other background subtracted cal.fits files in each filter, which is then scaled to the individual cal.fits files and subtracted. The super background subtracted cal.fits files are first corrected for row/column striping, then have residual background removed using \texttt{Background2D}, after which they are used to create the final Stage 3 mosaic images. With the background subtracted from each image, the astrometry is corrected by matching to Cosmic Assembly Near-infrared Deep Extragalactic Legacy Survey \citep[CANDELS,][]{Grogin_2011, Koekemoer_2011}  imaging \citep[as described in more detail in][]{bagley2023} prior to stage 3 processing in the pipeline. This results in the final science images, weight maps, and uncertainty images with a pixel scale of 0$\farcs$09/pix  registered to the CANDELS v1.9 \textit{HST}/F160W images.\par 

\begin{deluxetable*}{cccccccccccc}[t]
\tablecaption{Summary of our Final Sample and MIRI fluxes and 3$\sigma$ upper limits on \spitzer/MIPS and \herschel/PACS \label{tab:sample}}
\tablewidth{0pt}
\tablehead{
\colhead{CEERS} & \colhead{RUBIES} & \colhead{R.A. [deg]} & \colhead{Decl. [deg]} & \colhead{$z_{spec}$\tablenotemark{$\dagger$}}  &\colhead{f$_{F770W}$} &\colhead{f$_{F1000W}$}  &\colhead{f$_{F1500W}$}  &\colhead{f$_{F2100W}$} &\colhead{f$_{24 \mu m}$} &\colhead{f$_{100\mu m}$} &\colhead{f$_{160 \mu m}$}  \\ 
\colhead{ID} & \colhead{ID} & \colhead{(J2000)} & \colhead{(J2000)} 
}
\startdata
3153 & \nodata & 214.925762 & 52.945661 & 5.09  & 0.15{$\pm$}0.04 & 0.11{$\pm$}0.07 & 0.31{$\pm$}0.21 & 1.10{$\pm$}0.64 & {$<$}14.06 & {$<$}10.63 & {$<$}17.13 \\[0.13cm]
10444 &  49140 & 214.892248 & 52.877406 & 6.69  & 1.40{$\pm$}0.04 & 1.40{$\pm$}0.06 & 2.60{$\pm$}0.16 & 1.20{$\pm$}0.56 & {$<$}13.49 & {$<$}10.60 & {$<$}17.25  \\[0.13cm]
13135 & 42232 & 214.886801 & 52.855376 & 4.95  & 0.41{$\pm$}0.03 & 0.45{$\pm$}0.06& 0.42{$\pm$}0.16& 0.38{$\pm$}0.50 & {$<$}13.78 & {$<$}10.48 & {$<$}16.39  \\[0.13cm]
13318 & 42046 & 214.795367 & 52.788848 & 5.28  & 1.50{$\pm$}0.04 & 1.30{$\pm$}0.08 &2.10{$\pm$}0.27 &2.60{$\pm$}0.94 & {$<$}13.70 & {$<$}11.05 & {$<$} 17.01 \\[0.13cm]
2520 & 980841& 214.844768 & 52.892101 & 8.69  & 0.26{$\pm$}0.04 &0.48{$\pm$}0.07 & 0.28{$\pm$}0.21 & \nodata & {$<$}13.98 & {$<$}10.56 & {$<$}16.59 \\[0.13cm]
20496 & 927271 & 215.078257 & 52.948504 & 6.79  & 0.14{$\pm$}0.04 & 0.21{$\pm$}0.07 & 0.19{$\pm$}0.20 & 0.64{$\pm$}0.62 & {$<$}14.06 & {$<$}10.69 & {$<$}17.43 \\[0.13cm]
24253 & \nodata & 214.97996 & 52.861078 & 6.23  & 0.88{$\pm$}0.05 & 0.83{$\pm$}0.07 & 0.80{$\pm$}0.20 &1.50{$\pm$}0.68 & {$<$}14.77 & {$<$}10.28 & {$<$}16.58 \\[0.13cm]
20320 & \nodata & 214.876032 & 52.806109 & 5.79  & 0.32{$\pm$}0.03 & 0.43{$\pm$}0.06  & 0.51{$\pm$}0.16 & 0.10{$\pm$}0.48 & {$<$}14.03 & {$<$}10.79 & {$<$}16.29 \\
\hline \hline
\enddata
\tablecomments{We present fluxes here in units of $\mu$Jy with the exception of the \herschel/PACS data, which are reported in mJy.}
\tablenotetext{\dagger}{Spectroscopic redshifts are measured from the NIRSpec prism data for all sources except CEERS 24253 and 20320, which are measured from the NIRCam grism data.}
\end{deluxetable*}

The MIRI photometry is measured using \textsc{T-PHOT} \citep[v2.0, ][]{Merlin_2016}, where our priors are taken from the CEERS NIRCam catalogs for consistency with our parent sample \citep[][see Section~\ref{sec:sample}]{kocevski2024}. \textsc{T-PHOT} uses priors from the \jwst/NIRCam F277W for the lower resolution MIRI images for photometric analysis. The point spread function (PSF) for each MIRI band was taken from \cite{Libralato_2024}, which generated Nyquist sampled ePSFs (effective PSF) models and geometric-distortion corrections for MIRI. The kernels are then constructed to match the PSF from the \jwst/NIRCam F277W image (FWHM of \(\simeq\) 0.092")) to the MIRI images (FWHM of \(\simeq\) $0.2\arcsec - 0.5\arcsec$)) to extract source photometry with \textsc{T-PHOT}. To estimate uncertainties on object fluxes, an rms map was constructed for each MIRI mosaic to account for pixel-correlated noise. The fluxes from \textsc{T-PHOT} and their uncertainties estimated with these rms maps for each source are used as the MIRI catalog shown in Table~\ref{tab:sample}.
\par

\subsection{NIRCam Imaging and Photometric Catalog}
Our NIRCam catalog was created from the CEERS images, which were processed using the \textsc{JWST Calibration Pipeline} \citep[v1.8.5;][]{Bushouse2022} with custom modifications as described in \cite{Finkelstein2022} and \cite{bagley2023}. The resulting images were aligned to the CANDELS v1.9 \textit{HST}/F160W images and combined into a single mosaic for each field using the drizzle algorithm with an inverse variance map weighting \citep{Fruchter2002, Casertano2000}. The Resample step in the pipeline was used leaving the final mosaics with pixel scales of 0.03"/pixel. Source detection and photometry on the NIRCam mosaics were computed on PSF-matched images using SExtractor \citep{Bertin_1996} version 2.25.0 in two-image mode, with an inverse-variance weighted combination of the PSF matched F277W and F356W images as the detection image. Photometry was measured in all of the available NIRCam bands in each field, as well as the F606W and F814W HST bands using public data from the CANDELS and 3D-HST surveys \citep{Brammer2012, Momcheva2016}.\par

\subsection{NIRSpec Spectroscopy}
The NIRSpec prism/G395M observations used in this study are part of the RUBIES program (GO-4233; PI: A. de Graaff). However, the prism data for one source, CEERS 3153, was obtained from the DDT program 2750; PI: P. Arrabal Haro. Below, we briefly summarize both programs, their data reduction processes, and the methods used to obtain our 2D and 1D spectra. We refer the reader to \cite{deGraaff2024} and \cite{ArrabalHaro2023} for a full description. \par 

The RUBIES observations were taken with the low-resolution (R$\sim$100) prism/Clear and the medium-resolution (R$\sim$1000) G395M/F290LP modes. The data were reduced using the \textsc{JWST Calibration Pipeline} \citep[v1.13.4;][]{2024Bushouse13.4} with the Calibration Reference Data System (CRDS). The uncalibrated exposures are processed through the Stage 1 steps of the pipeline, which are then treated for large cosmic-ray snowball events \citep{Rigby2023} before the ramp-fit step. A 1/f correction is then applied where the 2D spectra is then flat-fielded and flux-calibrated. The sky background is then removed by taking the differences of the 2D spectra obtained at the three nod offsets. The three offset exposures are then combined and the wavelength grids are fixed for all prism and G395M spectra with sampling close
to the that of the native detector pixels. A cross-dispersion profile for each source in the frame of the curved 2D traces in the detector cutouts with parameters for a spatial offset and a scalar Gaussian width that is added in quadrature to a Gaussian approximation to the wavelength-dependent PSF. The rectified 2D spectrum is then optimally extracted, which gives the 1D spectrum. \par

The DDT program 2750 was configured to observe 160 faint-object targets, where also a three-point nodding pattern was employed to facilitate background subtraction. The NIRSpec data were processed with the \textsc{JWST Calibration Pipeline} (v1.8.5) and the CRDS. The calwebb detector1 pipeline module was employed to subtract the bias and dark current, correct the 1/f noise, and generate count-rate maps (CRMs) from the uncalibrated images. Similar to above, the parameters of the jump step are modified for a correction of ``snowball'' events, which are then processed with the calwebb spec2 pipeline module. The individual nods images are combined at the calwebb spec3 pipeline stage, making use of customized apertures for the extraction of the one-dimensional (1D) spectrum. Data from the three consecutive exposure sequences are combined after masking image artifacts to produce the final 2D and 1D spectral products.\par

\subsection{Far-Infrared Imaging and Catalog}
For completeness, we measure the 3$\sigma$ upper limits on \spitzer/MIPS 24 \mmicron, \herschel/PACS 100 and 160 \mmicron\ for the LRDs in our sample, though in practice none are detected. At the location of each source, we measured the flux density and errors using \texttt{photutils} \citep{Bradley2016}, adopting a forced photometry methodology with position priors from F277W. For our \spitzer/MIPS 24 \mmicron\ image mosaics in the EGS, we use the Far-Infrared Deep Extragalactic Legacy Survey \citep{Dickinson2007}, and for the \textit{Hershel}/PACS mosaics we use HerMES \citep{Oliver2012} public data release (DR4). We include our measurements of the upper limits in Table~\ref{tab:sample}.

\begin{figure*}[t]
\centering
\includegraphics[scale=0.45]{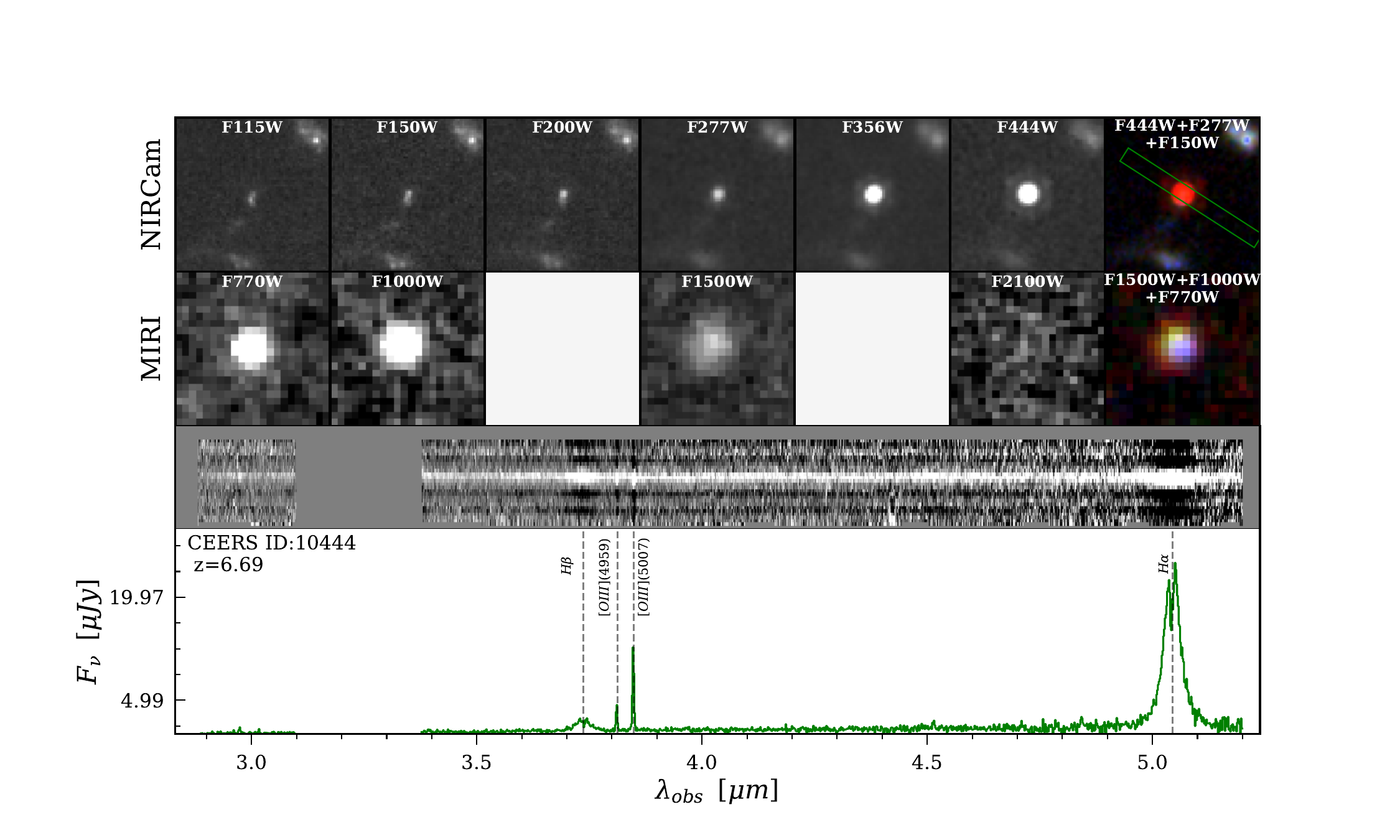}
\caption{\jwst/NIRCam (top row) and MIRI (second row) of 10444. Each image is $2\arcsec \times 2\arcsec$.  The last images in each row are the RGB images with 
F444W (R) + F277W (G) + F150W (B) for NIRCam, and F1500W (R) + F1000W 
(G) + F770W (B) for MIRI. In the NIRCam RGB image, we provide 
approximate NIRSpec slit positions with the green rectangle. The bottom 
two rows show the 2D and 1D NIRSpec G395M data, with prominent emission lines labeled.}
\label{fig:sampleshow}
\end{figure*}

\section{Sample} \label{sec:sample}
\subsection{Selection Criteria}
The parent sample for our work are the photometrically selected LRDs in the EGS from \citet[hereafter K24]{kocevski2024}. The definition of an LRD is taken from  \cite{Barro2024} and \cite{labbe2023}, which define LRDs as sources with a compact morphology that are red at rest-frame optical wavelengths and blue in the rest-frame UV, with our specific selection described here in more detail. K24 identifies LRDs using the UV and optical continuum slopes, defined as $f_\lambda \sim \lambda^\beta$ (with $\beta _{UV}$ and $\beta _{opt}$, respectively), measured by fitting photometry in multiple bands blueward and redward of the Balmer break at 3645\AA. The red sources with a UV-excess are then selected using an optical continuum slope cut of $\beta _{opt}$ $>$ 0 and a UV slope cut of  $\beta _{UV}$ $<$ $-0.37$. These criterion correspond to color cuts of F277W$-$F444W $>$ 1.0 mag and F150W$-$F200W $<$ 0.5 for sources at z $\sim$ 5. To remove brown dwarfs from the selection, a cut of $\beta _{UV}$ $< -2.8$ is applied. Lastly, an additional cut of  F277W$-$F356W $> -1$ and F277W$-$F410M $>$ $-1$  is used to identify sources whose $\beta _{opt}$ might be boosted due to strong emission lines affecting one or more bands. Given that one of our sources, CEERS 2520, has $z_{spec} >$ 8, only the second condition is imposed. Our parent sample from K24 consists of 66 LRDs. \par

We then select LRDs detected in the MEGA MIRI imaging with S/N > 3 in the F770W band, which we selected after visually confirming that LRD S/N typically decreases in the redder MIRI bands. This leaves 20 LRDs from our parent sample. Next, we select the LRDs with spectroscopic coverage from \jwst, leaving eight LRDs. Six of these objects have coverage from NIRSpec from the CEERS and RUBIES programs (see above). The remaining two objects (CEERS ID: 20320 and 24253) have spectra from the CEERS NIRCam/grism data with clear detections of the \hb\ + \oiii\ emission lines, but with no continuum detected. Because of the latter, we cannot use these spectra for our spectrophotometric fitting (see below), but we will fit their photometry with models after fixing the redshifts of the objects. We show the \jwst/NIRCam, MIRI, and NIRSpec G395M data of one object in our main sample, CEERS 10444, in Figure~\ref{fig:sampleshow} with the rest of the sample presented in Appendix~\ref{sec:postage_stamps_cont}. Additionally, we provide a summary of the sample in Table \ref{tab:sample}, with the measured MIRI fluxes in units of $\mu$Jy. \par

\begin{figure*}
    \centering
    \includegraphics[scale=0.55]{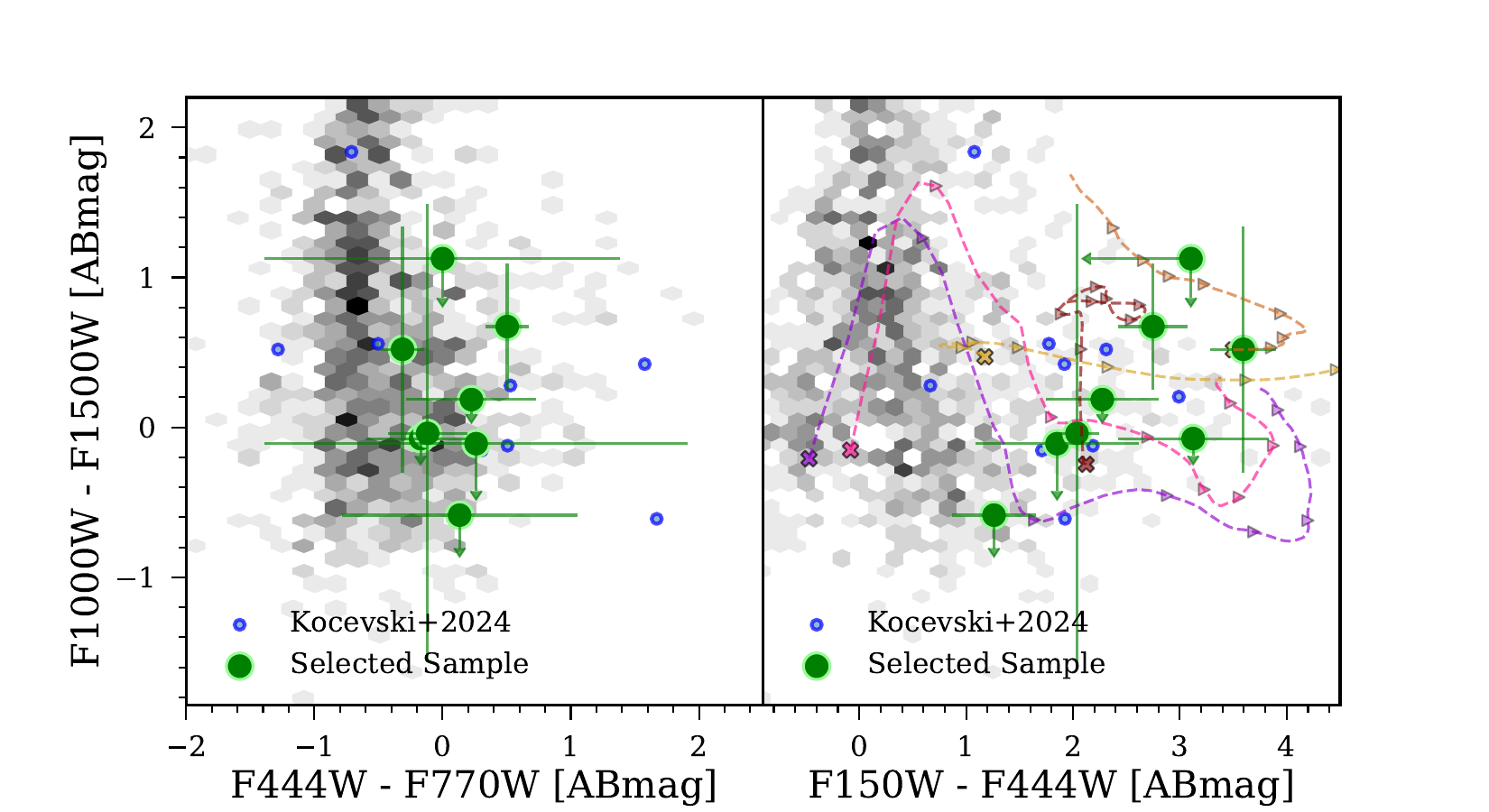}
    \caption{(Left) MIRI color F1000W $-$ F1500W compared to the NIRCam--MIRI F444W $-$ F7770W color for our sample (green), our parent sample from K24 (blue), and the full MEGA dataset (grey hexbins). (Right)  MIRI color F1000W-F1500W compared to NIRCam F150W $-$ F444W colors.   We have added tracks to show the redshift evolution in MIRI color of  SWIRE SED models \citep{Polletta2007} for a Type 2 QSO (yellow), Torus (light brown), Seyfert 2 (pink), Sb Star Forming Galaxy (purple), and Mrk231 (maroon).  The ``$\mathbf{x}$'' symbol marks $z=0.5$, with  each subsequent triangle showing integer redshdift from $z=1$ to 7.  These models broadly span the observed range in the LRD colors.}
    \label{fig:samplemiricolor}
\end{figure*}
\subsection{Sample Properties \label{subsec:sample_desc}} 
Figure \ref{fig:samplemiricolor} presents the NIRCam---MIRI colors for both our parent and selected samples. By selection, the LRDs in our sample are all red in NIRCam F150W -- F444W $>$ 1.  However, they show flat NIRCam -- MIRI colors of $-0.3 < \mathrm{F444W - F770W} < 0.5$ where the MIRI colors cover a range of $-0.6 < \mathrm{F1000W - F1500W} < 1.1$. These broad range of MIRI colors for the LRDs suggests a diverse emission origin. The left panel of Figure~\ref{fig:samplemiricolor} includes expected colors for different types of reference templates from \citet{Polletta2007} at redshifts $0.5 < z < 7$, including a type 2 QSO, a dusty torus, a Seyfert 2, a star-forming galaxy, and Mrk 231. These roughly span the range of colors seen in the LRDs samples, but imply that the LRDs may have a combination of star-formation and AGN contributing to their observed near-IR/mid-IR emission. Interestingly, four sources have significant (SNR $>3$) detections up to F1500W (CEERS 10444, 13318, 20320, and 24253), with no one source having a significant detection in F2100W. \par

The spectra of the LRDs show noteworthy features. There are two LRDs with visible Balmer breaks (CEERS 10444 and 13318) as well as four LRDs that show broad line emission (CEERS 10444, 13318, 3153, and 13135) visible in their \textit{NIRSpec}/G395M and/or prism data, as shown in Figure~\ref{fig:sampleshow} and in Appendix~\ref{sec:postage_stamps_cont}. For each LRD, we measure the line flux and FWHM of each line which we report in Table~\ref{tab:linefits}. Additionally, we measure the  strength of the Balmer breaks reported in Table~\ref{tab:breakmeas}. A detailed description of our methodology for measuring these properties is provided in Appendix~\ref{sec:linefit}. \par 

Lastly, we estimate the sizes of the LRDs in our sample in the NIRCam and MIRI imaging.   We fit the effective radii of our objects using \textsc{galfit} \citep{Peng_2002,Peng_2010} in \jwst/NIRCam F444W and \jwst/MIRI F770W following \cite{Shen2023}. Most of our objects are consistent with being point sources in the F444W data.  The exceptions are CEERS 10444, 24253, and 2520, which are spatially resolved. These sources have sizes similar to those reported in \citep{Baggen_2024, guia2024} of 131$\pm$8 pc and 123$\pm$3 for CEERS 10444 and 24253, and 323$\pm$10 pc for CEERS 2520. In the MIRI F770W data, all of our sources were unresolved with the exception of CEERS 10444 and 13318, with r$_e$ of 505$\pm$34 pc and  513$\pm$28 pc, respectively. It is noteworthy that CEERS 10444 is resolved in both NIRCam and MIRI, which suggests extended emission most likely from star formation even in the presence of an AGN (see below). In contrast, CEERS 133318 is resolved in MIRI but unresolved in NIRCam F444W, implying a more compact stellar or AGN component relative to the more extended near-IR emission. Interestingly, these sources also have the strongest Balmer breaks among our sample. We will discuss these objects further in depth below in Sections~\ref{subsec:alt_models}. \par

\section{Methods} \label{sec:methods}
We jointly fit the multi--wavelength photometry and NIRSpec/prism data using two different SED modeling codes, \prospector\ and \cigale. The photometry includes the \jwst\ NIRCam and MIRI photometry, and limits from \spitzer/MIPS and \herschel/PACS.  We use these two codes to test how differing assumptions and templates built into the framework of the codes will impact the best-fit models and their interpretations for the physical behavior of LRDs, in particular how the models treat AGN.  Both codes have been updated to fit the \textit{NIRSpec}/prism data, however the methodology by which the prism data is treated differs between the codes, which we discuss below in Sections \ref{subsec:prospy_sed_fit} and \ref{subsec:cigale_sed_fit}. \par

With each code we test three SED modeling cases motivated from LRD observations: (1) a star-forming (SF) only model, where the SED contains the contributions of stellar populations only; (2) an AGN--dominated model, where the AGN accounts for at minimum 50\% of the emission at rest-frame 5500\AA; and (3) a composite model where the AGN contribution and SF parameters are free. To allow for a fair comparison between the codes we maintain as similar parameter spaces as possible.  However, the codes have differences in stellar libraries, treatment of the star formation histories (SFH), AGN templates, treatment of emission lines, and requirements on energy balance between the UV/optical and IR. These differences in the codes are useful as they allow us to examine how these systematic factors shape our understanding of LRDs, and how well these models reproduce the observed data.  \par 

\subsection{SED fitting with \prospector} \label{subsec:prospy_sed_fit}
\prospector\ is an SED modeling code that inferences parameter estimates using nested sampling of the posterior parameter distribution.  This allows it to consider complex models and fully explore the parameter spaces.  We use a modified version of \prospector\ 1.4.0\footnote{See \href{https://github.com/bd-j/prospector/tree/60bc8195692d9965cd943f9e1caa750241b7ef57}{this link} for \prospector\ v 1.4} for this work. \prospector\ samples the posteriors on parameters using \dynesty\ \citep[][]{speagle2020}, which is a dynamic nested sampling method which has the benefits of Markov chain Monte Carlo (MCMC) algorithms, but is able to estimate and sample complex multi-modal distributions. Our fitting generally follows the methodology discussed in \cite{wang2023, wang2024}.  \par 

\prospector\ allows the simultaneous modeling of low-resolution spectra, such as those from {NIRSpec}/prism, and photometric data. However, before running \prospector, we took an additional step to correct for any flux offsets between the {NIRSpec} spectra and photometry, which could arise from wavelength-dependent slit losses, uncertain slit locations of the object within the slit, or other aperture corrections.  For this step, we scale the spectra to match the photometry by multiplying the spectrum with a 2nd order polynomial, which is fit to minimize a $\chi^2$ statistic.  
We then use the NIRSpec/prism instrumental resolution curve provided on the \jwst\ User Documentation\footnote{Please see 
\href{https://jwst-docs.stsci.edu/jwst-near-infrared-spectrograph/nirspec-instrumentation/nirspec-dispersers-and-filters\#gsc.tab=0}{this link} for more} and determine the dispersion by which we need to broaden the Medium-resolution Isaac Newton Telescope \citep[MILES,][]{Sanchez2006} stellar spectral libraries, taking into account the instrumental and model dispersion. 

We scale the uncertainties of the photometry to account for systematics in aperture corrections or 
in the flux calibration, both of which  will scale with the measured flux density 
\citep{Papovich_2001,Noll2009}. We therefore add, in quadrature, 10\% of the  flux density to the measured error on each photometric data point.  Likewise, we add, in quadrature, 10\% of the flux to the  uncertainties of the prism data to account for any systematics that would scale with the flux density. Both of these steps are consistent with the error treatment enacted by \cigale\ (see below).  \par

The emission lines need special treatment in the fitting.  We modified \prospector\ to fit a two component Gaussian model in the case of broad emission lines, corresponding to a narrow and a broad component.  For the other lines, we use the  default setup of the \prospector\ \texttt{nebular\_marginalization} module, which models the nebular and AGN emission lines as a single component Gaussian \citep{Johnson2021}. As discussed in Section~\ref{subsec:sample_desc} and shown in Figure~\ref{fig:sampleshow} and Appendix~\ref{sec:postage_stamps_cont}, at least five of the LRDs in our sample show evidence of having one or more broad-line components in the emission lines even at the resolution of the prism spectra.  Following \citet{wang2024}, we model the Balmer emission lines together, matching the redshifts and line widths of the multiple components.  Similarly, we model the line widths of the other narrow (metal) lines.  With these modifications we use three parameters to fit each emission line: \texttt{sigma\_metal} for the metal lines and \texttt{sigma\_broad} and \texttt{sigma\_narrow} for the Hydrogen lines. \par

For the stellar, nebular, and dust components of our model, we use the MIST stellar isochrones \citep{Choi2016, Dotter2016} with the MILES stellar spectral library in FSPS \citep{Conroy2010} and a Chabrier \citep{Chabrier2003} IMF. We implement a flexible SFH that models the SFR as the mass formed in seven logarithmically--spaced time bins \citep{Leja_2019} where the width of the first bin is decreased to 13.47 Myr \citep{wang2024}. For the dust attenuation we use \citep[][i.e., \texttt{dust\_type=4}]{Kriek2013}, which models the diffuse dust using a \citep{Calzetti2000} attenuation curve and implements the power-law slope from \citep{Noll2009}. Lastly, the dust emission uses the model of \cite{Draine2007}.\par 

For the AGN component of our \prospector\ models, we implement the continuum emission from the AGN accretion disk of  \cite{Temple2021} into \prospector\ v1.4. This model covers the rest-frame UV/optical and is characterized by a broken power law. We fix the slopes of best fit power law reported in Table 3 from \cite{Temple2021}, where the normalization of the AGN continuum is a free parameter, described as the ratio of the AGN to stellar flux densities at rest-frame 5500\AA. We assume that the AGN continuum emission is attenuated by the dust attenuation of the stellar population with the possibility of additional dust attenuation using the dust attenuation curve described by \citet{Kriek2013}. We also allow for  the AGN torus model implemented in FSPS \citep{Conroy2009, Leja2018}, which is based on a \texttt{CLUMPY} model of \cite{Nenkova2008}. The torus is modeled with with two free parameters, the normalization and dust optical depth in the mid-IR. We note that \prospector\ does not enforce an energy balance between the amount of light absorbed by dust from the AGN continuum and reemitted by the torus (this is a key distinction from \cigale, see below), which we discuss in more depth in Section~\ref{subsec:agn_results} and \ref{subsec:poss_agn}. We summarize the relevant parameters for \prospector\ listed in Appendix~\ref{sec:sedparams}. \par

\subsection{SED fitting with \cigale} \label{subsec:cigale_sed_fit}

\textsc{CIGALE} \citep{Boquien_2019} uses parametric SFHs to build composite stellar populations from simple stellar populations. The code calculates nebular emission produced by the stellar population.  It also generates thermal dust emission that balances the amount of starlight attenuated by dust, using flexible attenuation curves.    \cigale\ then  fits a large grid of models  to the data where the likelihood of each physical property is estimated by marginalizing over all other parameters. In contrast to \prospector, \cigale\ imposes an strict energy balance on the SED for the star forming \textit{and} AGN emission (as apposed to the stellar component only in \prospector).\par

We use a modified version of \cigale\ for our SED fitting, which allows for simultaneous fitting of the NIRSpec prism data and photometry, as recently discussed in \citet{burgarella2025}. This version of \cigale\ takes as input the prism resolution and then creates ''synthetic photometry`` points at the resolution of the spectra.  For this purpose, we again use the prism data that we scaled to match the photometry (see above). Given the NIRSpec/prism resolution ranges from 80 to 300 over the wavelength range, this corresponds to an additional 440 synthetic photometry points used in the modeling. As a result of this approach, \cigale\ cannot model the emission lines directly like \prospector. \cigale\ also does not have the ability to model broad emission features, so we mask those regions of the prism data to prevent the code from overestimating nebular emission.\par

For the stellar and nebular components we use the stellar population models of \cite{Bruzual2003}, the nebular-emission templates from \cite{Inoue2011}, which were generated using \texttt{CLOUDY} 13.01 \citep{Ferland1998, Ferland2013}, and a \cite{Chabrier2003} IMF. The SFH is described as a delayed $\tau$ model with periodic bursts. The attenuation is modeled using the \texttt{ dustatt\_modified\_starburst} module, which is similar to attenuation curve used in \prospector. Lastly, we use the \texttt{dl2014} module for the dust emission based on \cite{Draine2014}.\par 

For the AGN components in \cigale, we use the \texttt{SKIRTOR} module. Here, the rest-frame UV/optical emission from the AGN accretion disk is modeled using \cite{Schartmann2005}, which is parameterized as a broken power law. We adopt the \texttt{SKIRTOR} clumpy torus model \citep{Stalevski2012, Stalevski2016}, where the half-opening angle is fixed to the default 40$^{\circ}$ as it is observationally preferred \citep[][]{Yang2022}. The normalization of the AGN component (\texttt{agn\_frac}) varies from 0.5 to 0.9 for the AGN dominated models, and 0.1 to 0.9 for our composite models. We summarize the relevant parameters for \cigale\ in Appendix~\ref{sec:sedparams}.

\subsection{Bayesian Evidence for the Models}
After fitting the models to the data, we compare the results using the Bayesian information criterion (BIC). The BIC provides an estimate of the Bayes factor \citep{Raftery_1995}, and is intended for providing evidence against one model in favor of another. The BIC includes the likelihood function, and a penalty term for the number of parameters in each model.  Assuming that the model uncertainties are independent and normally distributed we use,
\begin{equation}\label{eqn:bic}
    \mathrm{BIC} = \chi ^2\ +\ k\ \ln(n), 
\end{equation}
where $n$ is the number of data points in our observations, $k$ is the number of parameters, and $\chi^2$ is defined below.  Here, we use the difference of BIC models $>$10 to be indicative of ``strong'' evidence to disfavor one model compared to another model \citep{Raftery_1995}.  We consider models with a difference in their BIC values $\leq 10$ to be too close to disfavor either model. For our purposes, we consider the star-forming model as the ``null'' hypothesis, and we will test if adding the AGN components produces $\Delta$BIC $>$ 10 such that we would reject the null hypothesis in favor of the AGN model. \par

For \cigale, $n$ is defined as the sum of the input photometric data and the binned prism data, while $k$ is the number of free parameters. For \prospector, $n$ is defined as the sum of the input photometric data and the length of the input prism array, while $k$ is the sum of the lengths of the parameter array ($\theta$).  Given the different treatments by \cigale\ and \prospector\ of the NIRSpec prism data (see above), they have different a number of data points, $n$, which it makes it difficult to compare the models directly. We will therefore refrain from doing so, but instead consider the outputs form both models as a test for systematics between the models.\par

Our calculation for the $\chi^2$ in Equation \ref{eqn:bic} considers both measurements and upper limits on our photometric data from \spitzer\ and \herschel. We use the derivation of the $\chi^2$ from Equation~14 and 15 from \cite{Boquien_2019}, where $\chi^2$ is calculated as,
\begin{equation}
    \begin{aligned}
        \chi^2 &\equiv \sum \frac{(f_{model} - f_{data})^2}{\sigma_{data}^2}\ \  - \\
        &2\times \sum \ln \left( \frac{1}{2} \left[ 1 +  \mathrm{erf} \left(\frac{(f_{model} - f_{data, upper})^2}{\sigma_{data, upper}^2} \right) \right] \right).
    \end{aligned}
\end{equation}
We then use the above to calculate the BIC using Equation~\ref{eqn:bic} for each set of models for each galaxy.   
\newcommand{\tableagn}[1]{\textcolor{red}{#1}}
\newcommand{\tablemaybe}[1]{\textcolor{orange}{#1}}
\newcommand{\tableno}[1]{\textcolor{blue}{#1}}

\begin{deluxetable*}{c|ccc|ccc|c}
\tablecaption{BIC Values Derived from SED fits \label{tab:bic}}
\tablewidth{0pt}
\tablehead{
\colhead{} &
\multicolumn{3}{c}{\cigale} & \multicolumn{3}{c}{\prospector} \\
\colhead{CEERS ID} &
\colhead{Star-forming} & \colhead{AGN-dominated} &
\colhead{Composite~~~~} & \colhead{~~~~Star-forming} & \colhead{AGN-dominated} & \colhead{composite} & \colhead{Interpretation}
}
\startdata
\multicolumn{8}{c}{Galaxies with NIRSpec/prism data} \\\hline
{3153} & 1070 & 745 & \tableagn{709} & 984 & \tableagn{872} & \tableagn{872} & \tableagn{AGN}\\
{10444} & \tablemaybe{1120} & 1200 & 1350 & 2060 & \tablemaybe{1300} & \tablemaybe{1300} & \tablemaybe{divided AGN evidence}\\
{13135} & 677 & 680 & \tableagn{660} & 1070 & \tableagn{1007} & \tableagn{1007} &  \tableagn{AGN}\\
{13318} & \tablemaybe{922} & {932} & 938 & 1140 & \tablemaybe{950} & 984 & \tablemaybe{divided AGN evidence} \\
2520 & 757 & \tablemaybe{707} & \tablemaybe{708} & \tablemaybe{927} & 950 & 952 &  \tablemaybe{divided AGN evidence} \\
20496 & 720 & \tablemaybe{638} & \tablemaybe{638} & \tablemaybe{1001} & 1005 & 1003 & \tablemaybe{divided AGN evidence} \\[6pt]\hline\hline
\multicolumn{8}{c}{Galaxies with NIRCam/grism data} \\\hline
{24253} & \tableno{75} & 82 & {67} & \tableno{82} & 97 & {95} & \tableno{no AGN evidence} \\
{20320} & \tableno{67} & {67} & {63} & \tableno{75} & {82} & {80} & \tableno{no AGN evidence} \\[6pt]
\hline \hline
\enddata
\tablecomments{The BIC values presented here are derived using equation~\ref{eqn:bic} for each LRD for the \cigale\ and \prospector\ results. We assume the LRD emissions can be reproduced with star-forming stellar populations only unless there is strong evidence (i.e., $\Delta\mathrm{BIC} \ge 10$) to prefer a more complex model with AGN components \citep{Raftery_1995}. The ``Star-forming'' models assume only stellar populations produce the SED, the ``AGN-dominated'' models require an AGN to produce $>$50\% of the light at 5500~\AA, and the ``Composite'' models allow for combination stellar populations and AGN with no restrictions. The last column provides the favored interpretation based on the BIC values. Readers should exercise caution when viewing this column, as it is not meant to suggest the physical processes of these LRDs. The color-coding shows BIC with evidence for ``AGN'' (red), ``divided AGN evidence'' (orange), and ``no AGN evidence'' (blue), see text.}
\end{deluxetable*}

\subsection{Emission Line Decomposition}
In additional the SED modeling, we fit the emission lines independently to provide direct measures of available line ratios, fluxes, and FWHM of all features present in the spectra of our objects, in addition to any broad emission features present. We measure these from the NIRSpec G395M if available and prism data otherwise. For the broad lines, we use a two component Gaussian with a narrow component restricted to have width $\sigma$ $<$ 350 km~s$^{-1}$, and a broad component restricted to have width $\sigma$  $>$ 350  km~s$^{-1}$ similar to K24. The other lines are modeled with $\sigma$ $<$ 350  km~s$^{-1}$.  In all cases we model the continuum under the emission lines as a 2nd order polynomial. Similar to our fits in \prospector, we fit the metal lines separately from the hydrogen lines. Lastly there are two objects, 10444 and 13318, which show a strong absorption feature in the Balmer emission lines in the medium-resolution spectra (see Figure~\ref{fig:sampleshow} and Appendix~\ref{sec:postage_stamps_cont}). In these cases, we include an additional Gaussian component to model this absorption feature, which qualitatively improves the overall fit of the lines.   In all cases, we use LMFIT \citep{Newville2024} to model the emission lines, which performs non-linear optimization using the optimization methods of \textsc{scipy.optimize} \citep{Virtanen2020} the Levenberg-Marquardt method. The results of the line fitting are presented Appendix~\ref{sec:linefit}. 

\section{Results} \label{sec:results}
\begin{figure*}
\centering
\includegraphics[scale=0.33]{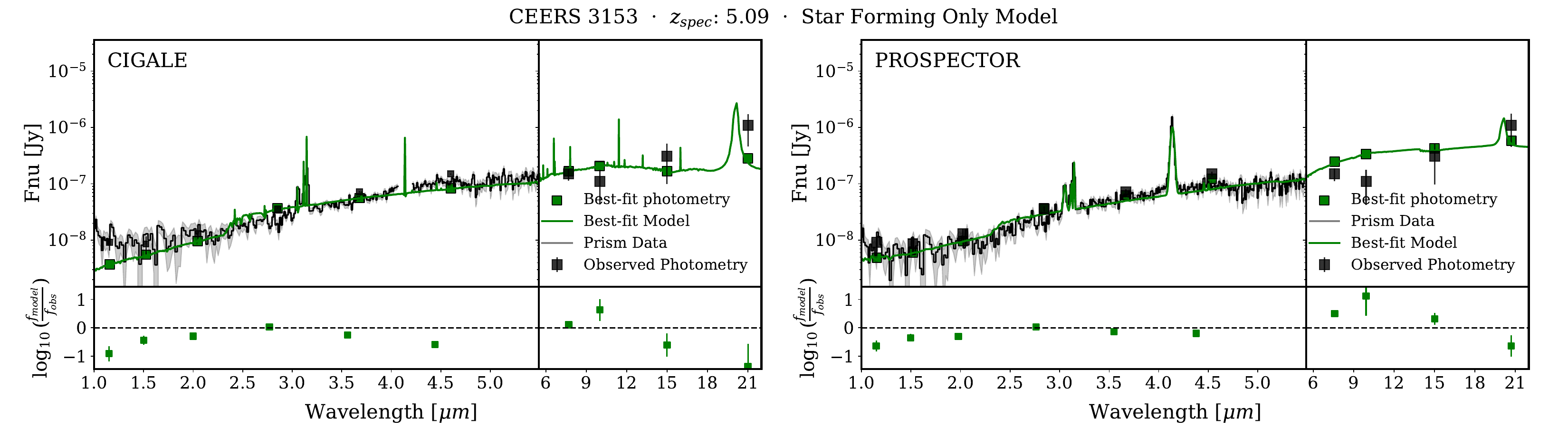}
\includegraphics[scale=0.33]{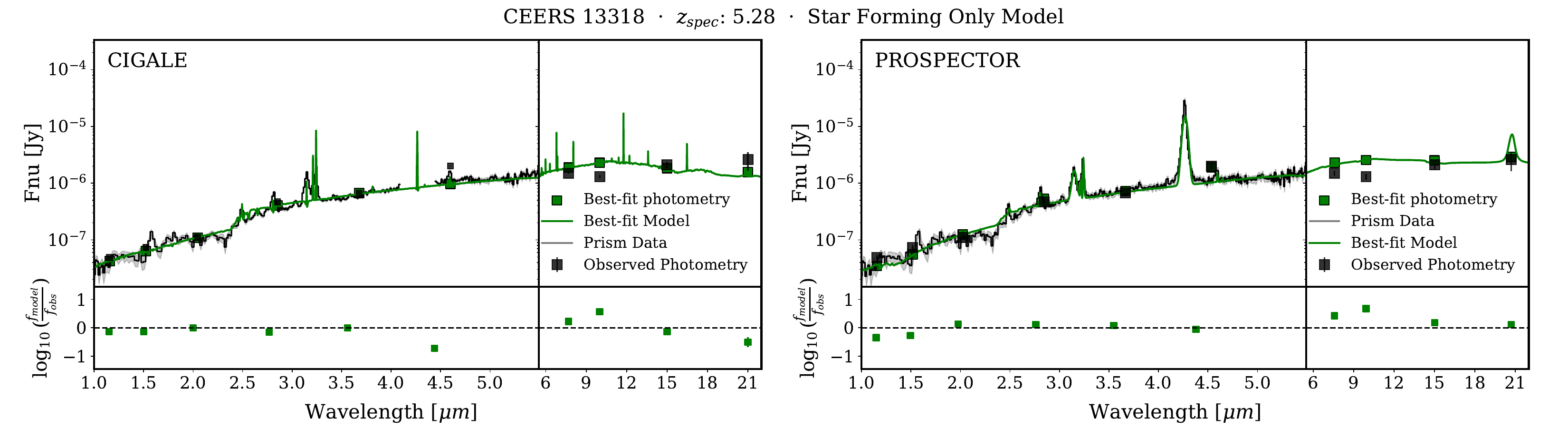}
\includegraphics[scale=0.33]{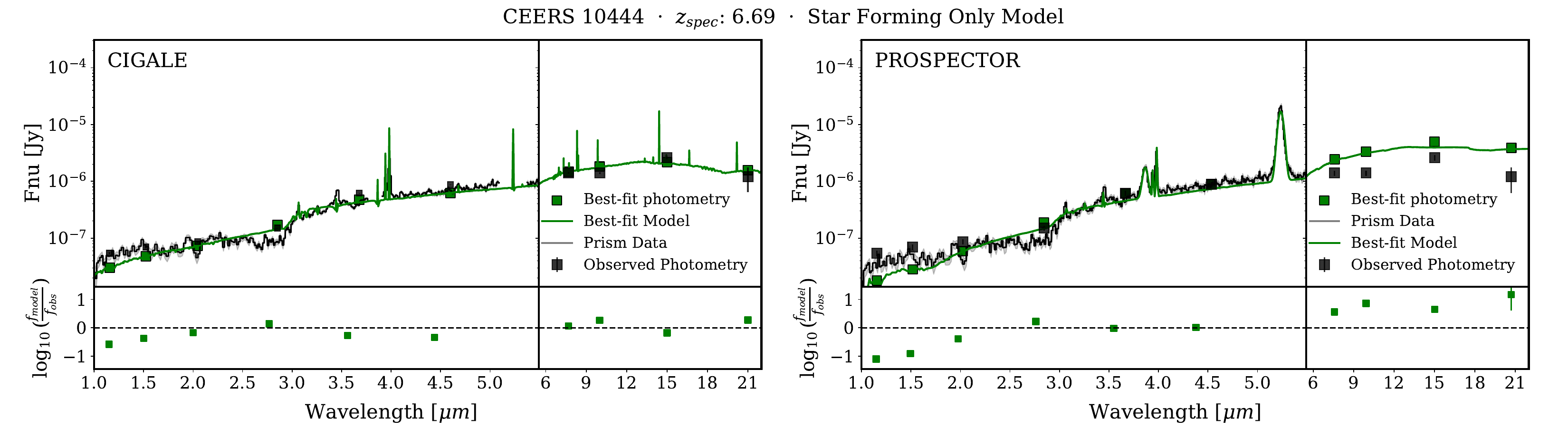}
\includegraphics[scale=0.33]{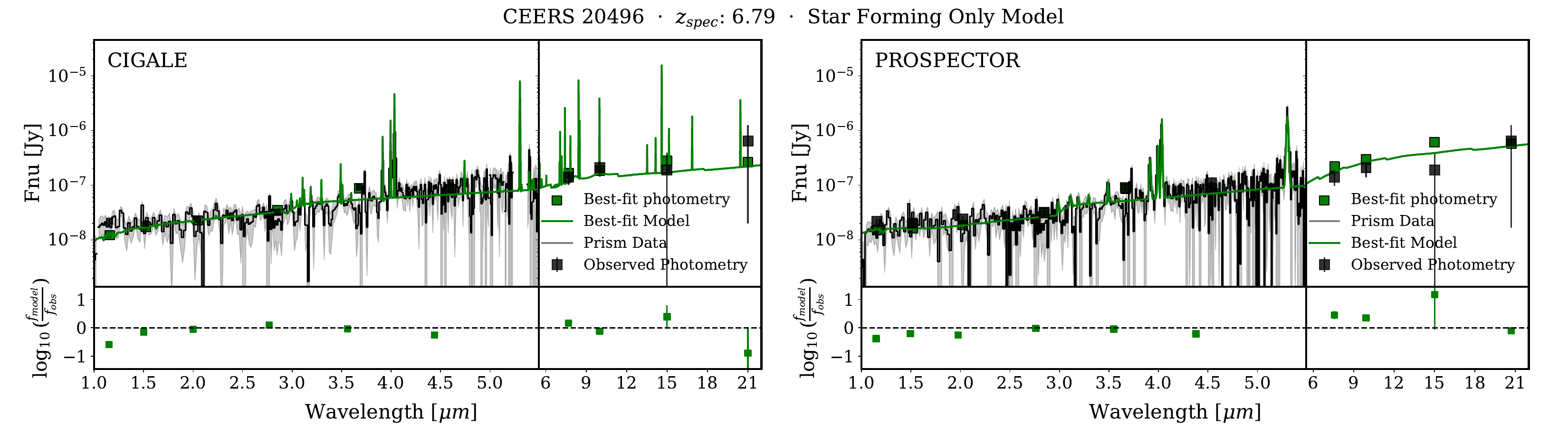}
\caption{SED fitting results for the star-forming–only models for four LRDs in our sample are presented in Figure~\ref{fig:sf_only_appendix} (results for the remaining galaxies are provided in Figure~\ref{fig:sf_only_appendix} in Appendix~\ref{sec:sedparams}). Each row corresponds to a single galaxy, identified by the ID number in the row title. The left panel of each row shows the best-fit SED from \cigale, while the right panel displays the best-fit SED from \prospector. \label{fig:sf_only}}

\end{figure*}

\begin{figure*}
\centering
\includegraphics[scale=0.33]{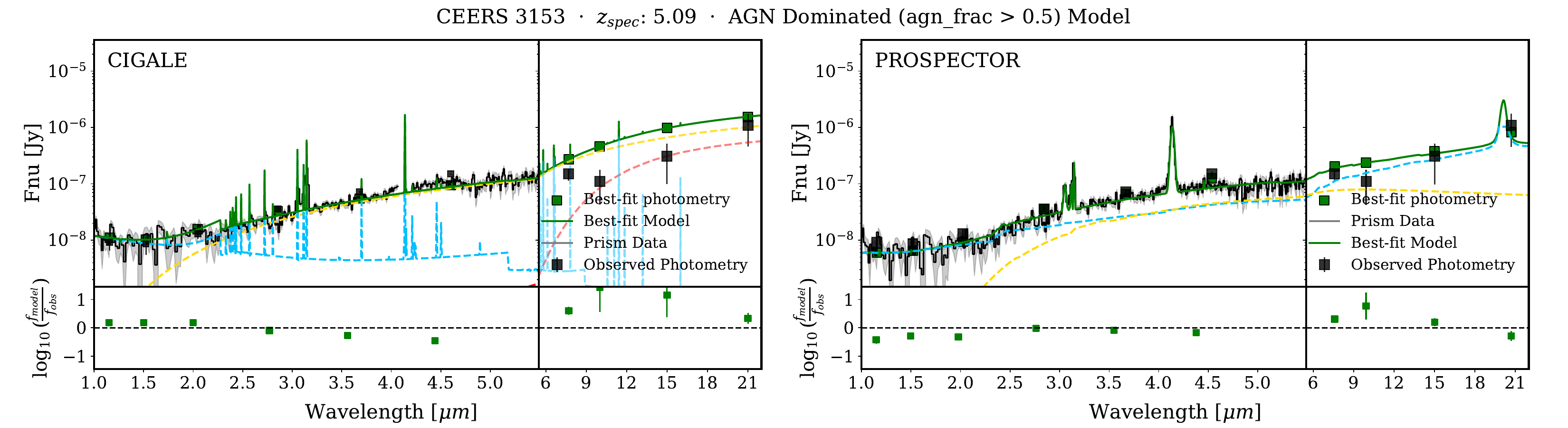}
\includegraphics[scale=0.33]{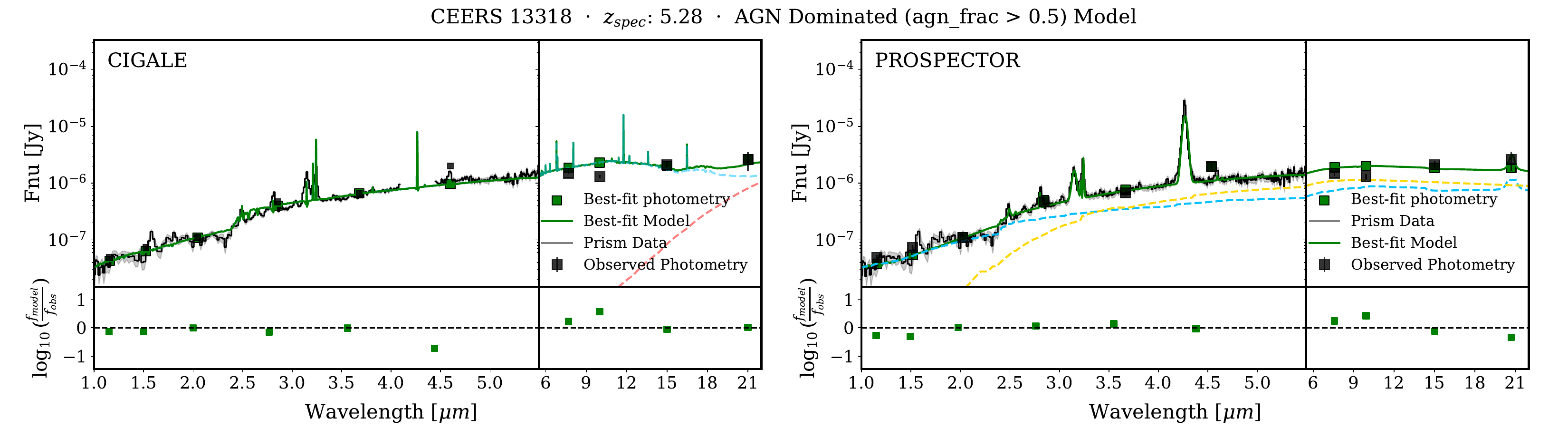}
\includegraphics[scale=0.33]{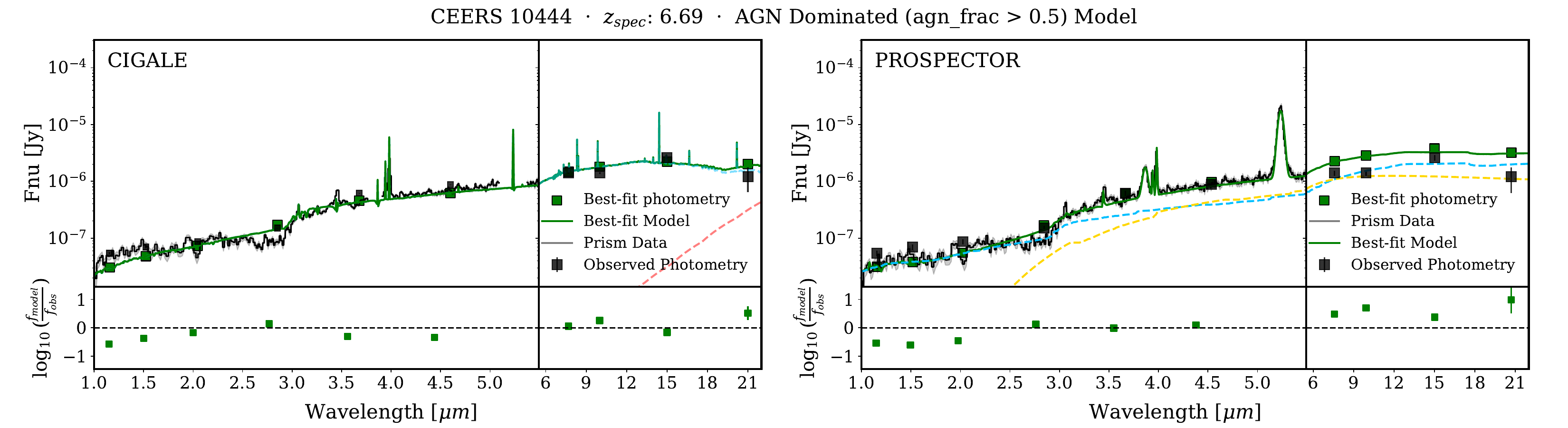}
\includegraphics[scale=0.33]{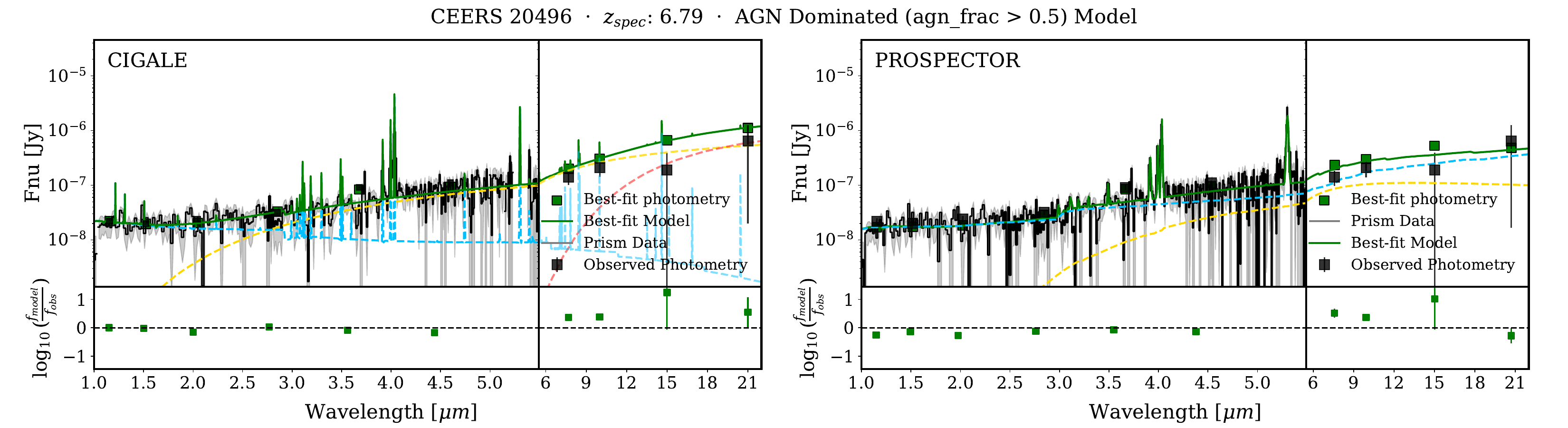}

\caption{SED fitting results for the AGN-dominated models for four LRDs in our sample (the remaining are shown in Figure~\ref{fig:agn_appendix} of Appendix~\ref{sec:sedparams}).  Each row shows the results for one galaxy, indicated by the ID number. In each row, the left column shows the best-fit SED results from \cigale\ the right column shows the best-fit SED for \prospector.  In each panel, the star-forming component is shown in blue.  The model for the AGN accretion disk is in yellow and the AGN torus component is red.   \label{fig:forceagn}}
\end{figure*}

In this Section we examine the results from our SED modeling methods from \cigale\ and \prospector.
We present results for the cases of (1) star-forming only, (2) AGN-dominated, and (3) composite AGN and star-forming models in Figures~\ref{fig:sf_only}, \ref{fig:forceagn}, and \ref{fig:composite} for CEERS 3153, 13318, 10444, and 20496. The rest of the sample results are provided in Appendix~\ref{sec:sedparams} in Figures~\ref{fig:sf_only_appendix}, \ref{fig:agn_appendix}, and \ref{fig:comp_appendix}. As a reminder, the difference between ``AGN dominated'' and  ``composite'' models is that the former allow \texttt{agn\_frac} to be at minimum 0.5 for both codes, while for the latter, \texttt{agn\_frac} is free to range from 0.1 to 0.9.\par 

To select the preferred model, we apply Equation~\ref{eqn:bic} to compute the BIC for each LRD for our modeling scenarios; the resulting values are shown in Table~\ref{tab:bic}. The final column of Table~\ref{tab:bic} shows our interpretation for each LRD, which is based solely on the BIC results from the SED models and guided by the following criteria:
\begin{itemize}
        \item ``no AGN evidence'' indicates that neither the \cigale\ nor \prospector\ fits yield sufficient improvement in the BIC ($\Delta$BIC $\leq$10) to prefer the more complex models with AGN over the star-forming model. 
        \item ``AGN'' means that the AGN-dominated and/or Composite models from \prospector\ \textit{and} \cigale\ yield sufficient improvement with $\Delta$BIC $>$10, which provides strong evidence in favor of an AGN from both codes. 
        \item ``divided AGN evidence'' means that either the \cigale\ or \prospector\ results (but not both) yield sufficient improvement with $\Delta$BIC $>$10, which provides strong evidence in favor of an AGN, but from only one of the codes.
\end{itemize}

We discuss the LRDs which BIC values from the modeling supports ``no AGN evidence'' in Section~\ref{subsec:no_agn_ev}, the LRDs where the BIC values which support ``AGN'' in Section~\ref{subsec:agn_results}, and the LRDs where the BIC values which suggest ``divided AGN evidence'' in Section~\ref{subsec:poss_agn}. Lastly, we discuss the results from our emission line fits and our Balmer Break measurements in Section~\ref{subsec:lineratios}.\par


\begin{figure*}
\centering
\includegraphics[scale=0.33]{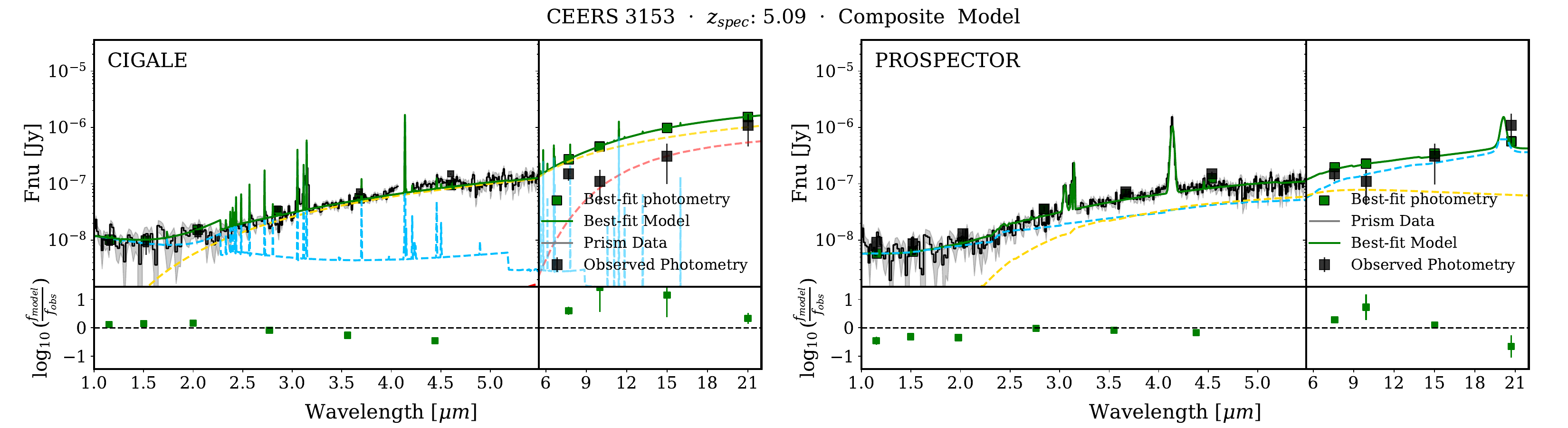}
\includegraphics[scale=0.33]{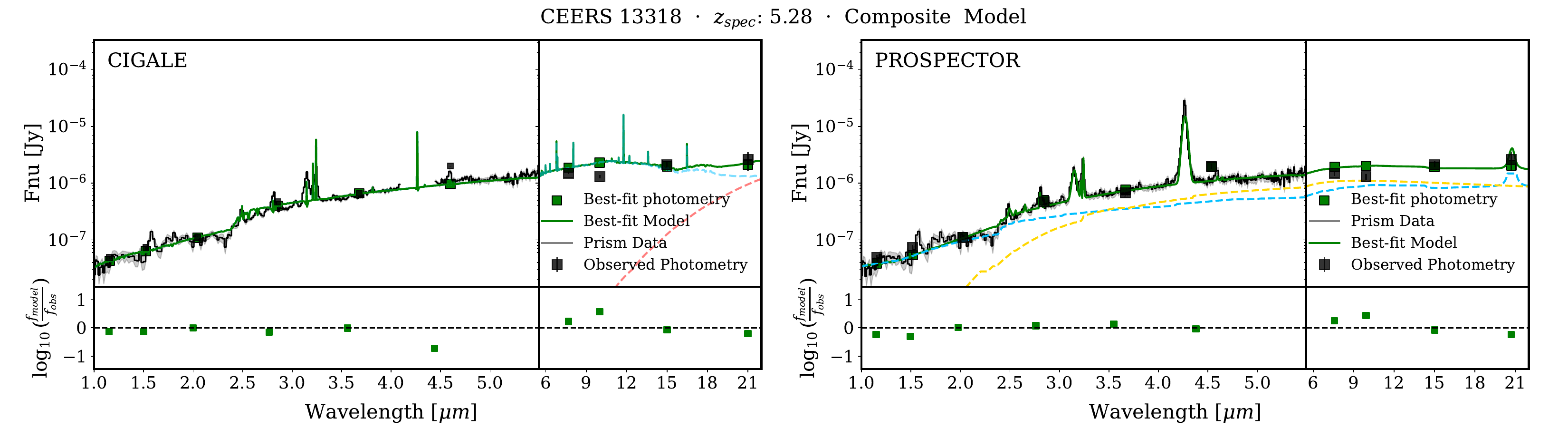}
\includegraphics[scale=0.33]{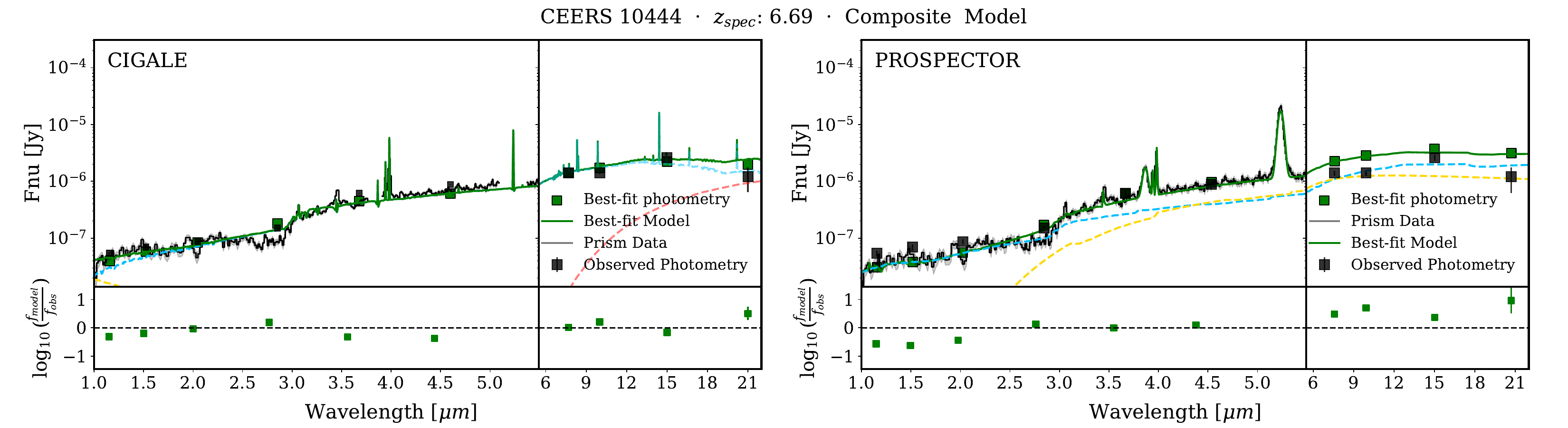}
\includegraphics[scale=0.33]{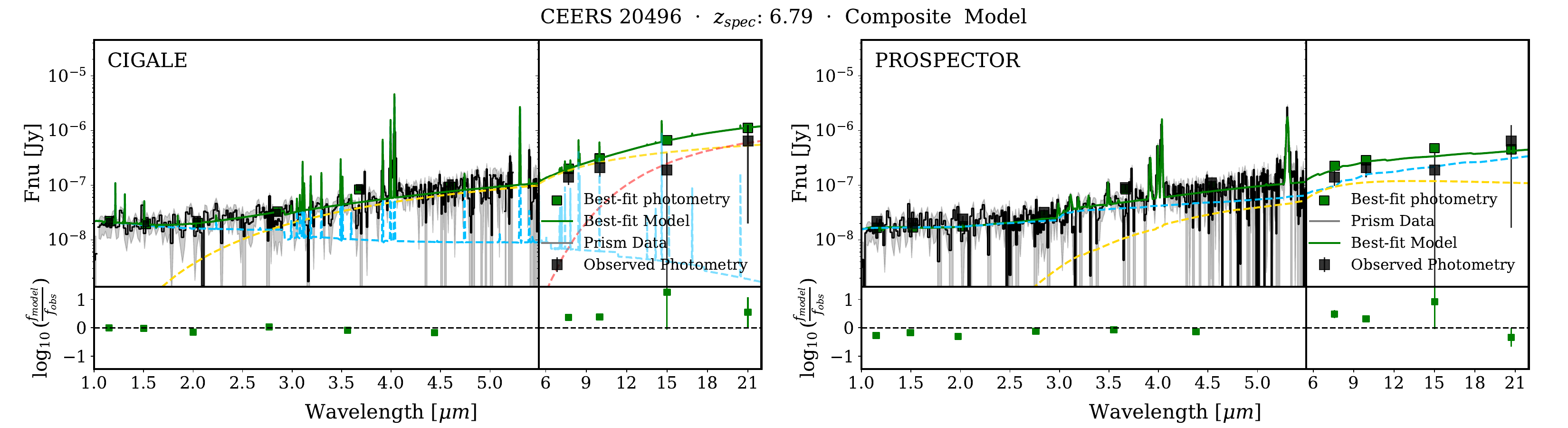}

\caption{SED fitting results for the composite AGN and star-forming models for four LRDs in our sample (the remaining are shown in Figure~\ref{fig:comp_appendix} of Appendix~\ref{sec:sedparams}).  Each row shows the results for one galaxy, indicated by the ID number. In each row, the left column shows the best-fit SED results from \cigale\ the right column shows the best-fit SED for \prospector.  In each panel, the star-forming component is shown in blue.  The model for the AGN accretion disk is in yellow and the AGN torus component is red.   \label{fig:composite}}
\end{figure*}

\subsection{LRDs with no AGN evidence}\label{subsec:no_agn_ev}

Based on the analysis of the SED fitting, two of the LRDs in our sample, CEERS 24253 and 20320, have ``no evidence of AGN'' (see Table~\ref{tab:bic}). For these two LRDs, the addition of the AGN components in both \cigale\ and \prospector\ do not improve the quality of the fit sufficiently ($\Delta$BIC $<$10) compared to the star-forming only models. Based on this evidence, we would favor the star-forming models over the AGN models. \par

Interestingly, CEERS 20320 and 24253 are the only two objects in our sample lacking NIRSpec/prism coverage; their redshifts, are instead derived using NIRCam grism data. Of the two, only CEERS 24253 is best fit by the star-forming model from both \cigale\ and \prospector, while CEERS 20320 favors the composite model from \cigale. However, in the latter case, the improvement is marginal ($\Delta$BIC < 10), providing insufficient evidence to reject the star-forming model. These results suggest that the addition of the AGN models in the codes do not improve the $\chi^2$ sufficiently compared to the number of additional parameters added, rather than confirm that these LRDs are purely star forming. These results may underscore the importance of incorporating NIRSpec prism data into SED modeling to robustly evaluate the AGN contribution in LRDs.

\subsection{LRDs with AGN evidence}\label{subsec:agn_results}

The \cigale\ and \prospector\ fits favor models that include an AGN component (either the ``AGN-dominated'' or ``Composite'' models) for CEERS 3153 and 13135. The improvement in the BIC values from the AGN models exceeds our threshold of $\Delta$BIC $>$10, which is considered as strong evidence to reject the star-forming models. \par

Among the AGN candidates identified by broad Balmer emission (see Section~\ref{subsec:lineratios}), only CEERS 3153 and 13135 are well-fit by AGN templates in both \prospector\ and \cigale. Inspecting the fits, the AGN components are needed to match the observed MIRI colors. In the case of CEERS 3153, the MIRI colors are the reddest of the sample, with $F1000W - F1500W = 1.1$~mag, and $F1500W - F2100W = 1.4$~mag. CEERS 13135 shows a downtown in the reddest MIRI band, with $F1000W - F1500W$ = $-$0.07 mag. \prospector\ models the observed colors of these objects with high AGN fractions (> 0.9). In contrast, \cigale\ assigns AGN fractions > 0.9 only to the reddest source, CEERS 3153.  The \cigale\ modeling of 13135 yields a more modest AGN fraction of 0.10. For these two objects the observed features (i.e., broad Balmer emission) and the MIRI data point to the presence of AGN in these two sources which can be well described by the AGN components in by both \cigale\ and \prospector. \par

A notable result is that in both cases the AGN models do not require a torus-dust component as implemented in either \prospector\ or \cigale. In \prospector, the observed-frame mid-IR emission is adequately modeled using only the AGN accretion disk and the host galaxy. In \cigale, the AGN best-fit models produce larger residuals in the MIRI bands, where the total flux overshoots the MIRI data as the torus contribution increases. Similar results were found for the bright LRD at z=3.1 in \cite{wang2024}. If indeed the MIRI emission of CEERS 3153 and 13135 stems from AGN, it does not need a traditional dust torus as implemented in \texttt{SKIRTOR} or \cite{Nenkova2008} for these LRDs. \par

\subsection{LRDs with divided AGN evidence} \label{subsec:poss_agn}

Four LRDs in our sample, CEERS 10444, 13318, 20496, and 2520, are categorized as ``divided AGN evidence'' in Table~\ref{tab:bic}. The modeling results of CEERS 10444 and 13318 shown in Figures~\ref{fig:sf_only}-\ref{fig:composite} show that both codes have difficulty simultaneously modeling the MIRI data and the shape of the rest-UV/optical continuum in the NIRSpec/prism data, particularly in the area around the Balmer break.  It is notable that these two objects have the the largest measured Balmer break indexes in our sample, with $BB = 2.88 \pm 0.23$ and 1.68 $\pm 0.16$, respectively. In the case of CEERS 10444, the Balmer break approaches the maximum allowed for a stellar population assuming a \citet{Chabrier2003} IMF. Additionally, these objects also have broad Balmer emission lines with FWHM $>2800$ kms$^{-1}$, which satisfies the AGN-selection criteria as reported in K24.  However, \cigale\ favors a dust-reddened star-forming model with a weak Balmer break (see, Figure~\ref{fig:sf_only}). \prospector\ favors the AGN model, where the rest-optical/near-IR portion of the SED is dominated by the AGN accretion disk and the rest-UV portion stems from star-formation.  Given that 10444 is resolved in F444W we can examine the implied densities derived from the star forming only model in \cigale\ and from the composite model in \prospector.  These give 4.33 $\log_{10}(M_\odot pc^{-3})$  and 2.49 $\log_{10}(M_\odot pc^{-3})$, respectively. Indeed, the added AGN component would remove tension with the high implied surface density from the modeling results. Taken together, the observational evidence for an AGN in these two sources and the inability of both codes to reproduce the Balmer breaks with a stellar population provide sufficient motivation to revisit these galaxies in Section~\ref{subsec:alt_models}, where we explore alternative AGN scenarios that may better reproduce the data. \par 

CEERS 2520 and 20496 are the two other LRDs in our sample in which the evidence from the BIC values from the \prospector\ and \cigale\ results disagree. Of the LRDs in our sample these two objects are among the faintest in both the NIRCam and MIRI bands, which may indicate there is insufficient signal-to-noise in the data to ``reject'' the null hypothesis of star-forming models only.   CEERS 2520 is the highest redshift object in our sample at $z_{spec}=8.69$ with no evidence of broad Balmer emission and a relatively weak Balmer break of 1.1$\pm 0.2$, and it is spatially resolved in the NIRCam imaging.  The implied densities inferred from the \prospector\ star-forming-only model (2 $\log_{10}(M_\odot, \mathrm{pc}^{-3})$) and \cigale\ AGN model (0.55 $\log_{10}(M_\odot, \mathrm{pc}^{-3})$). Both exceed typical value for the Solar Neighborhood \citep{Xiang2018} but fall on the lower end of the LRD distribution \citep{guia2024}. Interestingly, the detection of CEERS 2520 up to MIRI F1000W with SNR $\sim$6 at $z=8.7$ suggests a level of luminosity or stellar mass that might be difficult to achieve with purely stellar populations at such early epochs, hinting at the possibility of an AGN in this particular source. \par

The emission modeling of CEERS 20496 indicates the presence of a broad component in the Balmer emission lines, but this has the lowest FWHM of the sample, with FWHM $= 1710\pm 490$ km~s$^{-1}$, which is evidence favoring an AGN.   However, the \prospector\ fit favors the star-forming models, where the MIRI emission is predominantly produced by the star-forming component. In contrast, \cigale\ favors a model where the AGN component accounts for the MIRI emission.   It is noteworthy that the SNR of the MIRI F1500W data is $<1$, and is lowest of our sample (the F2100W data has only a slightly higher SNR). This may be a reason for the divided AGN evidence from \prospector\ and \cigale. 

\subsection{Spectral Line Analysis} \label{subsec:lineratios}

The fitted broad \ha\ components for CEERS 3153, 10444, 13318, and 20496 (Table~\ref{tab:linefits}) show FWHM values between 1700--3500 kms$^{-1}$, in contrast to the narrow \oiii\ lines which have FWHM of $\sim$300–-400 kms$^{-1}$. As discussed above, these observations are characteristic of AGN in the local Universe \citep[e.g.,][]{Harikane_2023} and are consistent with the findings of K24. However, objects CEERS 20496 and 13135 have broad \ha\ FWHM measurements that could also be consistent with compact blue dwarf star-forming galaxies \citep{Izotov2007}. These results are challenging to interpret within the framework of our preferred SED models, which in some cases can reproduce the data but omit physically relevant scenarios. This is made more evident with CEERS 10444 and 13318, which have Balmer absorption present in their G395M data, which are blue-shifted by 40 and 300~km~s$^{-1}$, respectively, that have FWHM = 570 and 770~km s$^{-1}$, respectively. These absorption features are not produced by any of the models yet considered here. \par

Motivated in particular by Balmer absorption in CEERS 10444 and 13318, coupled with growing interest in dense gas around an AGN as an origin for the strong Balmer breaks in LRDs, we explore the relationship between Balmer break strength and redshift for our LRDs. We use the definition of the Balmer Break strength from \citep{Wang2024_}, as the ratio of the average flux between rest-frame 4000\AA, 4100\AA)  and (3620\AA, 3700\AA).  This appraoch differs slightly from that of \cite{Balogh1999}, as the wavelength range is adjusted to mitigate contamination from possible strong emission lines. We report the results of our Balmer Break measurements in Appendix~\ref{sec:linefit}. The objects in our sample have Balmer Breaks that are consistent with the range of expectation assuming ``normal'' stellar population with a \cite{Chabrier2003} IMF.  While the Balmer Break strengths for our LRDs are less than that of other LRDs in the recent literature with extreme Balmer Breaks \citep[e.g.,][]{degraaff2025, naidu2025, taylor2025}, for at least two of our sources, CEERS 10444 and 13318, these Balmer breaks are not reproduced by any of the SED modeling results (see above). We will return to this point in Section~\ref{subsec:alt_models}. \par

\section{Discussion}\label{sec:discussion} 
The NIRSpec, NIRCam and MIRI data enable for a detailed multi-wavelength analysis with which to interpret the nature of LRDs. As we have demonstrated above, the LRDs in our sample show a mix of evidence from our modeling between being powered by AGN and star-formation. This suggests that these LRDs form a heterogeneous population with varying contributions from AGN and star formation to their SEDs. While both \cigale\ and \prospector\ are capable of modeling these mixed contributions, the AGN and star-formation prescriptions implemented in each code differ. The resulting discrepancies between the codes shown in Table~\ref{tab:bic} underscore the systematics inherent in SED modeling and highlight how these uncertainties propagate into the interpretation of LRDs.\par 

\begin{figure*}
    \centering
    \includegraphics[scale=0.6]{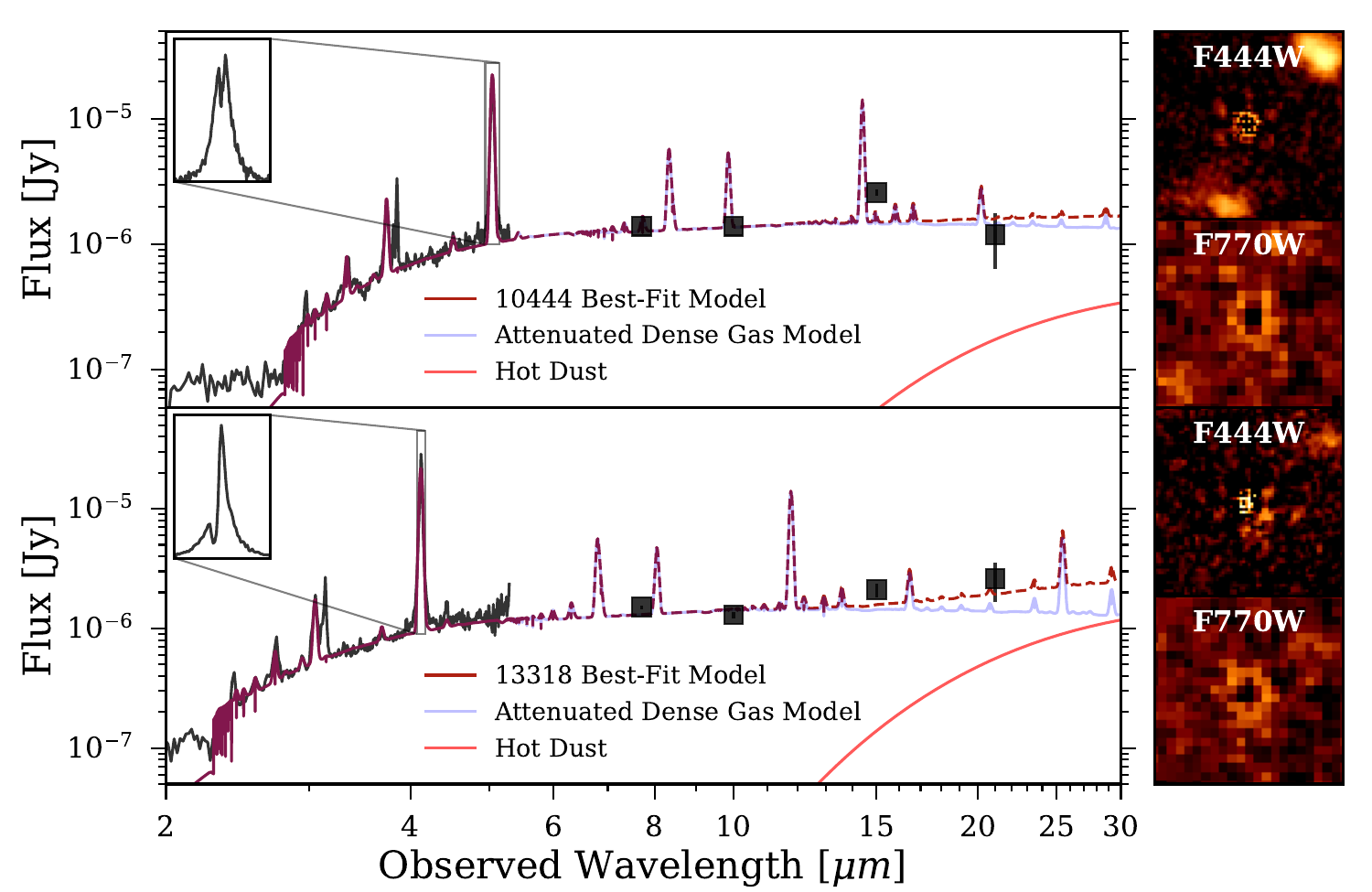}
    \caption{Results from our dense gas modeling for CEERS 10444 (top) and CEERS 13318 (bottom). The MIRI photometry and NIRSpec prism data are shown as the black points and curve, respectively. The inset in each panel shows a zoom in of the \ha\ feature at higher resolution with the NIRSpec/G395M data, which show the absorption superimposed on the emission line for both galaxies. The colored lines represent components of the best-fitting dense gas model: the dust-attenuated model excluding hot dust (lavender), the hot dust emission (red), and the total model (maroon). The maroon line is solid where it overlaps with the NIRSpec/prism data and dashed elsewhere. To the right of each SED, we show the NIRCam F444W and MIRI F770W images after subtracting a point source. The NIRCam residuals are consistent with zero but the MIRI residuals suggest extended emission that may indicate an underlying host galaxy.} 
    \label{fig:dense_gas_results}
\end{figure*}

Our modeling suggests that AGN are necessary to explain the SEDs of most of the LRDs in our sample, particularly those with NIRSpec/prism data (see Table~\ref{tab:bic}). However, while qualitatively \cigale\ and \prospector\ successfully reproduce the SED of many of the LRDs, both appear to be unable to match all the details in some LRDs, particularly in the region around the redshifted Balmer break.  It is therefore prudent to consider alternative AGN models as a viable explanation for the nature of these LRDs.  One alternative hypothesis is the “dense-gas” models, in which an accreting SMBH is surrounded by hot ($\log T/\mathrm{K} \gg 4$), dense ($\log n_\mathrm{H}/\mathrm{cm^{-3}} \gg 9$), Compton-thick gas \citep{Inayoshi_2025,degraaff2025,Naidu_2025,taylor2025}. In this section, we consider these models in Section~\ref{subsec:alt_models} to determine whether they can reproduce the NIRSpec and MIRI emission. We then return to modeling systematics in Section~\ref{subsec:systematics_models}, and we assess the implications for the nature of LRDs and their inferred bolometric luminosities from our modeling in Section~\ref{subsec:bol_lum}.\par

\subsection{Dense Gas Models}\label{subsec:alt_models}
As discussed above, two objects in our sample, CEERS 10444 and 13318, have prominent Balmer breaks observed in the NIRSpec prism data, which are not well reproduced from the SED models alone in \cigale\ nor \prospector.  Moreover, both CEERS 10444 and 13318 are the only two objects in our sample with observed blue-shifted Balmer absorption superimposed on the Balmer emission lines based in the G395M data (see Figure~\ref{fig:dense_gas_results}, inset panels).  These facts are evidence for high column densities of partially ionized gas. These features are found in other LRDs, and can be reproduced by dense, thick gas heated by an AGN without any stellar contributions \citep{Inayoshi_2025}. We therefore explore whether these dense-gas models can  reproduce the modeling of the SED for CEERS 10444 and 13318.

Following the formalism of \citet{taylor2025}, we consider a suite of dense, thick gas models generated using the \texttt{Cloudy} photoionization code. We adopt the default AGN spectrum, which is modeled as a broken power-law with $\alpha_X = -0.5$, $\alpha_\mathrm{UV} = -0.1$, where we allow $\alpha_{OX}$ to vary over $(-2.5, -2.0, -1.5)$.   We then consider the gas to have a fixed metallicity of [Fe/H] $= -2$, variable gas density $\log \left( n_\mathrm{H}/\mathrm{cm}^{3} \right) \in (9, 12)$ in increments of 0.5 dex, variable gas temperature $\log T/\mathrm{K} \in (4, 4.7, 5, 5.7)$; variable ionization parameter $\log U \in (-3.5, -0.5)$ in 0.5 dex steps, variable column density $\log \left( N/\mathrm{cm}^{2} \right) \in (21, 26)$ in 1 dex increments, and variable gas turbulence $v/\mathrm{km~s}^{-1} \in (100, 500)$ in steps of 100 km s$^{-1}$.  \texttt{Cloudy} then computes the radiative transfer of this input spectrum through this medium.  We allow for the nebular emission to have a non-unity covering factor,  $C_f \in (0,1)$, and we apply dust attenuation using the attenuation curve of \citet{Salim2018}. We then resample these models with the NIRSpec prism and integrate them with the MIRI filter curves before fitting them to the data.  \par 

Through experimentation, we found the attenuated dense gas model alone fails to reproduce both the NIRSpec prism data and MIRI photometry, particularly at the longer wavelengths. To reconcile this, we added a thermal dust component with temperature $T_\mathrm{dust} \in (500, 2000)$~K where the upper bound is set by the thermal sublimation temperature of dust \citep{Kobayashi2011}. This component is normalized by a factor from $(0,1)$ relative to the luminosity of the dense gas absorbed by the dust component above. We then fit these models to the NIRSpec prism and MIRI data using \dynesty\ \citep{speagle2020} to sample the posteriors for each variable.  In the fitting, we restrict the data to rest-frame wavelengths  $>3646$~\AA\ as the these models do not include adequate flux below the Balmer limit. The results of our modeling are shown in Figure~\ref{fig:dense_gas_results}. \par 

\begin{figure*}
    \centering
    \includegraphics[scale=0.5]{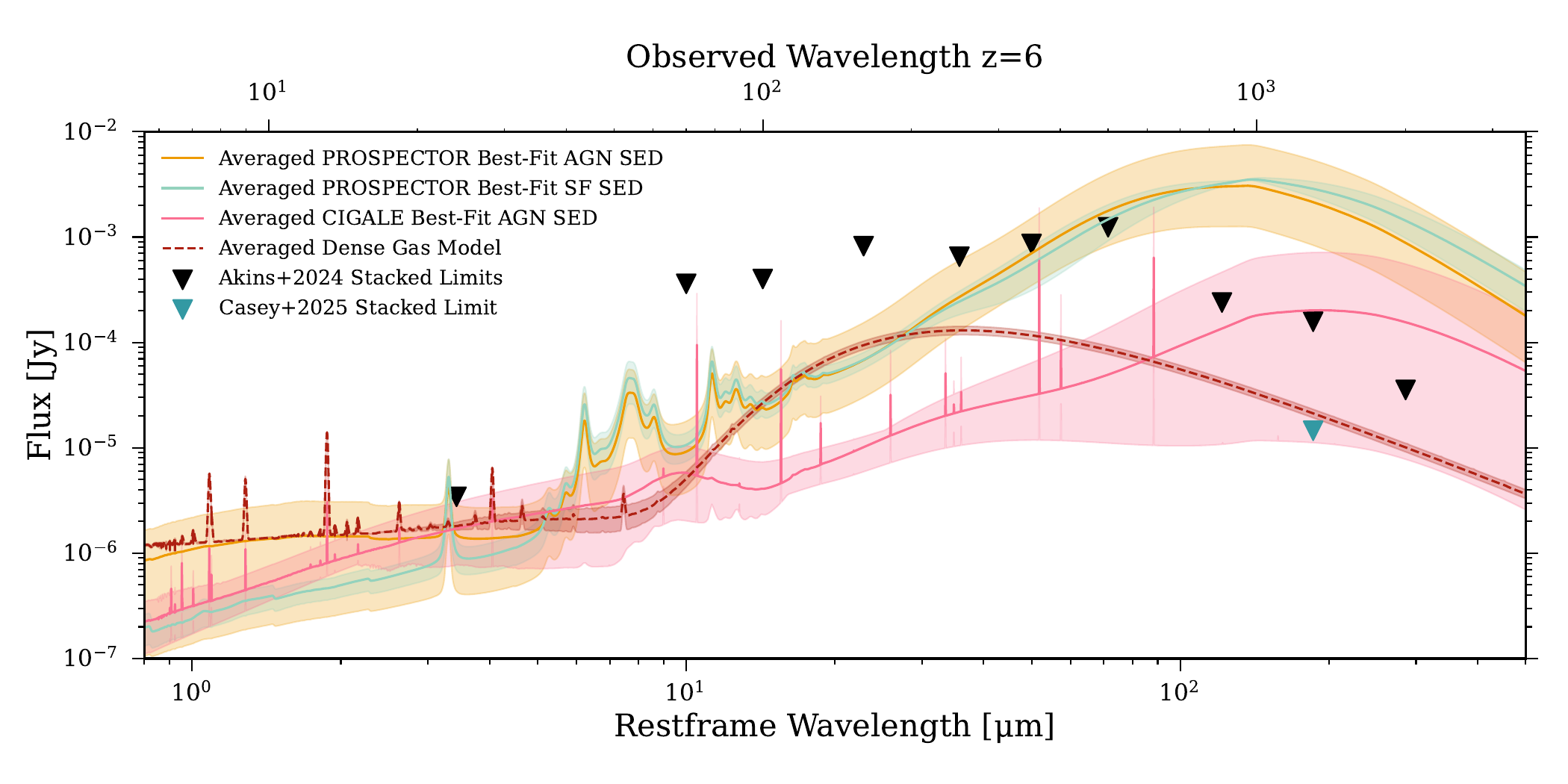}
    \caption{Averaged best-fit SEDs for the LRDs in our sample from \prospector\ and \cigale\ based on the BIC analysis in Table~\ref{tab:bic}. The averaged SEDs are shown as a function of rest-frame and observed frame ($z=6$, roughly the median redshift of the LRDs in our sample) wavelength shown in the top abscissa.  The \prospector\ star-forming (SF) averaged model includes CEERS 2520 and 20496 (blue).  The \prospector\ AGN model includes CEERS 3163, 10444, 13135, and 13318 (orange).  The \cigale\ AGN model includes CEERS 3153 and 13135, 2520, and 20496 (pink). Additionally, we include the averaged dense-gas model of CEERS 13318 and 10444 (maroon). The shaded region around each line corresponds to the minimum and maximum values of the  models contributing to each averaged SED. The black and green triangles are the stacked 5$\sigma$ upper limits from \citet{Akins_casey_2024} and \citet{Casey2025}, respectively.}
    \label{fig:bol_seds}
\end{figure*}

Our fits show that the dense-gas models reproduce the shape of the SED at wavelengths longer than rest-frame Balmer limit for CEERS 13318 and 10444. For these models, both galaxies favor high gas temperatures, $T\sim (1-5)\times 10^{5}$~K, high gas density, $\log \left( n_\mathrm H/\mathrm{cm^{3}} \right) \simeq 10.5-11$, and high column densities, $\log N_\mathrm{H}\simeq 24-26$. The latter result is intriguing as it implies the gas is Compton thick, which would account for the weak or absence of X-ray detection in LRDs \citep{Ananna_2024, Yue_2024}. The observed emission lines in both objects require a sub-unity covering factor ($C_f$) of 0.5 for CEERS 10444 and 0.75 for CEERS 13318. Lastly, the model requires moderate dust attenuation of $A_V = 0.52 \pm 0.03$~mag for CEERS 13318 and $0.47 \pm 0.02$~mag for CEERS 10444.  In both cases the dust attenuation is in agreement with the $A_V \leq 1$ upper limit proposed by \citet{chen2025_}, and (as we discuss below) consistent with the limits on the thermal emission from dust derived from \cite[and references therein]{Casey2025}. 

Given that the dense-gas models require some dust absorption, the attenuated light must be reemitted at longer wavelengths.  This may account for the rising hot-dust component required to reproduce the longer wavelength MIRI data.  In both CEERS 13318 and CEERS 10444, this hot-dust component has a temperature of $T_\mathrm{dust} = 600$--1300~K. However, this hot dust corresponds to only a fraction of the total light absorbed. For CEERS 13318, this component accounts for 30\% of the total emission absorbed by dust, and for CEERS 10444 it is only 10\% of the total.  
If the dense gas models are indeed correct, then there must be additional emission presumably from cooler dust at longer wavelengths (we will return to this point below).
\par

Though promising, these dense-gas models are insufficient to account for \textit{all} of the emission properties of our LRDs. First, CEERS 13318 and 10444 have strong, narrow \oiii\ $\lambda\lambda$4959, 5007 emission lines that are not replicable by this model. In the dense-gas models, the gas density exceeds the critical density for these lines ($\approx 7.0 \times 10^{5}$~cm$^{-3}$, \citealt{osterbrock1989}). We argue that this could be a either the signature of the host galaxy with nebular emission lines from star-formation, or possibly an additional narrow-line region of the AGN (see \citealt{lin2025}).  
Possible additional evidence for an underlying star-forming component comes from the fact that both CEERS 13318 and 10444 show possiblu extended emission in the MIRI F770W band, showing it could be spatially resolved (Figure~\ref{fig:dense_gas_results}). 
%
%
The extended emission is not seen in the NIRCam F444W data.  If the MIRI emission is sptially resolved, it would argue for incorporating additional emission presumably from star formation in addition to dense-gas models at rest-frame wavelengths beyond $\sim$ 1 \mmicron. Lastly, there remains some emission in the UV at wavelengths below the redshifted Balmer limit for which the dense-models can not explain. Thiss component could come from star-formation with the host galaxy, or possibly scattered light from the AGN \citep{labbe2023, lin2025}. To conclude, while these dense-gas models account for some of the LRD properties, they do not explain all of the observations. Further studies are needed to evaluate their relevance and applicability to the LRD population.   

\subsection{Modeling Insights from Averaged LRD SEDs}\label{subsec:systematics_models}

Here, we compare the average SEDs of our modeled LRDs to constraints measured from stacked samples in the far-IR and radio from \cite{Akins_casey_2024} and \cite{Casey2025}, and refer the reader to these works for further details on sample construction and methodology. Although none of the SED models considered here reproduce all the properties of all of the LRDs in our sample, we can evaluate if these models themselves are consistent with these longer wavelength constraints. We construct average SEDs for LRDs with NIRSpec/prism coverage (excluding CEERS 24253 and 20320), using the best-fit star-forming and AGN models from \prospector\ and \cigale\ based on the BIC values from Table~\ref{tab:bic}. Specifically, we averaged the following (best-fit) models using the nomenclature from Section~\ref{sec:methods} above: 
\begin{itemize}
    \item \prospector\ star-forming models for CEERS 2520 and 20496; 
    \item \prospector\ AGN-dominated models for CEERS 3153, 10444, 13135, and 13318; 
    \item \cigale\ AGN-dominated models for CEERS 2520, 20496, 3153, and 13135; 
    \item dense-gas models for CEERS 13318 and 10444.
\end{itemize}
%
For \cigale, we do not consider the star-forming models as this is favored only for CEERS 13318 and 10444, where the FWHM of the \ha\ emission line is too broad to be explained from star formation alone.  Rather, for these two galaxies we use the results from the dense-gas models in Section~\ref{subsec:alt_models}, with an addition of a cooler dust component (discussed in the next paragraph).   

For the average SED of the dense-gas models for CEERS 10444 and 13318 we extend the best-fit models to longer wavelengths,  For this purpose, we assume that the dust must emit the emission that it absorbs in a cooler dust component.  As a reminder, the modeling in Section~\ref{subsec:alt_models} finds a dust attenuation of $A(V) \simeq 0.5$~mag, assuming the \citet{Salim2018} dust law.   We therefore add a cool dust component where 100\% of the absorbed flux is emitted as a blackbody with a temperature of $T=140$~K, which is arguably appropriate for super-Eddington dust heated by an AGN \citep{McKinney2021}.  We further note that this assumption is conservative since the MIRI data show evidence that at least \textit{some} of this dust emission emerges in a component at higher temperatures ($\sim 600-1300$~K, see Section~\ref{subsec:alt_models}).   This makes this component at reasonable upper limit on the amount of light produced by this mode. 
%

 For each group above, we average the SEDs and show these results in Figure~\ref{fig:bol_seds} as a function of rest-frame wavelength, and shifted to an observed wavelength of $z=6$ to facilitate comparison to the far-IR and radio limits.  This redshift, $z=6$, also corresponds roughly to the median redshift of the LRDs in our sample.\par 
 
Figure~\ref{fig:bol_seds} shows that the average best-fit AGN models from \cigale\ overpredict the contribution from cold dust compared to the limits from stacking LRDs in the sub-mm and radio.    This is shown by the average model (solid pink line) and scatter in the models (shaded pink region) in Figure~\ref{fig:bol_seds}.  The average model is inconsistent with the upper limits at wavelength $>$1~mm.    The shaded region shows that the lowest luminosity model, which is set by one of our galaxies, CEERS 3153.  For this galaxy, its best-fit AGN model from \cigale\ is consistent with the stacked limits.  However, we note that the residuals for the MIRI data for CEERS 3153 are particularly large, which would indicate that the \cigale\ AGN model is overpredicting the contribution of hot dust due to the rising torus emission at those wavelengths (see Figure~\ref{fig:forceagn}).  Therefore, it may be that the \cigale\ model can account for the emission in at least \textit{some} of the LRDs without violating the stacked limits.  If this is the case, then this argues that some of the LRDs may  require the presence of a dusty torus as represented in the \texttt{SKIRTOR} model.  \par

The \prospector\ best-fit AGN and star-forming models overpredict the contribution from cold dust to the SED at wavelengths $\gtrsim$850~\micron, as illustrated by the yellow and green lines and curves in Figure~\ref{fig:bol_seds}. This similarity between the AGN and star-forming models at these wavelengths results because \prospector\ does not require significant AGN contribution beyond the rest frame UV/optical wavelengths.  As a result, the emission at longer wavelengths (e.g., the far-IR and radio) stems from dust associated with star-formation in both the star-forming and AGN models.   It therefore seems unlikely that these models represent the total emission well, at least for the majority of LRDs as these would violate the upper limits in the stacked data.  This is similar to the conclusion from \cite{Labbe2025} who argued the sub-mm limits for LRDs in their sample are inconsistent with the amount emission attenuated by dust in star-forming regions.\par 

The average dense-gas models appear to be consistent with limits from stacking in the literature, which is made evident from the dashed dark-red line and shaded region in Figure~\ref{fig:bol_seds}. This is in part due to us requiring the ``cold'' dust component to have $T = 140$~K, which forces the peak of the dust emission to be at shorter wavelengths. If we assumed a lower temperature for this component, the models would exceed the stacked limits from the literature.  Therefore, this model remains viable for LRDs as it would not violate the upper limits from stacking, but this prediction requires confirmation by measuring the shape of the galaxies' SED at far-IR and radio wavelengths.  \par 

To summarize, the best-fit models favor a scenario where LRDs are powered by an AGN.  This is required for the models to reproduce the NIRspec prism data, NIRCam and MIRI imaging, and stacked upper limits at far-IR and radio wavelengths. Particular AGN models such as the \texttt{SKIRTOR} templates in \cigale\ or the dense-gas model appear to be able to satisfy these constraints, but each alone does not appear to capture fully the range of features across our sample. In contrast, the star-forming models are able to fit the NIRSpec and NIRCam and MIRI imaging data but would require an excess of emission in the far-IR, sub-mm, and radio that is inconsistent with stacking from the literature.  \par 

We must acknowledge that there are some caveats to the the above conclusions. First, given the limited size of our sample, it remains unclear on how representative our sample is compared to those used in the stacking analyses. This in part due to our sample requirements of both NIRSpec prism coverage and detections in MIRI, which favor intrinsically brighter sources. Our results may be more appropriate for more luminous sources, which may have higher far-IR and sub-mm flux densities that the ``average'' reported in the stacking. Future work with modeling larger samples will be needed to test these caveats.\par

\subsection{Constraints on the Bolometric Luminosities and SMBH Growth timescales of LRDs}\label{subsec:bol_lum}

Finally, we are able to estimate the bolometric luminosities for the LRDs in our sample using the average SED models discussed above and shown in Figure~\ref{fig:bol_seds}.  By integrating these models, we place constraints on the total luminosity of our LRDs. This yields the following measurements in units of $\log_{10} \left( L_\mathrm{bol}/L_\odot \right)$ : 
\begin{itemize}
    \item \prospector\ star-forming model, 12.3$^{+0.10}_{-0.11}$; 
    \item \prospector\ AGN model, 12.3$^{+0.40}_{-0.40}$;
    \item \cigale\ AGN model, 11.3$^{+.0.40}_{-0.40}$;
    \item dense-gas model,  11.6$^{+0.07}_{-0.03}$. 
\end{itemize}

The \prospector\ averaged SF and AGN  models return high bolometric luminosities, which is unsurprising as the bulk of the emission in these models is obscured by dust and re-emitted in the far-IR. While these results are in agreement with the bolometric luminosity ranges from \cite{labbe2023}, \cite{Kokorev_2024}, and \cite{Akins_casey_2024}, it seems unlikely that LRDs have such high bolometric luminosities. Where we instead consider $\log\left(L_\mathrm{bol}/L_\odot\right) \sim12$ to be an upper limit for the LRDs in our sample.  \par 

The bolometric luminosity returned from the \cigale\ AGN models are lower, as these models differ in their treatment of the AGN obscuration. While these models require less emission in the far-IR, they over overestimate the flux in the MIRI bands. It seems unlikely that these models fully explain all the properties of the LRDs in our sample, even though they are at the limit of what we consider consistent with the far-IR, sub-mm and radio upper limits from the literature (particularly for CEERS 3153).\par    

The dense gas models favor lower bolometric luminosities, most likely attributable to the fact that these models do not require as dust attenuation, $A(V) \simeq 0.5$~mag, such that there is less emission required at longer wavelengths.   As a result, these models are potentially the only models that are consistent with the upper limits in the far-IR--through--radio. If these models are correct, then it might suggest that LRDs have bolometric luminosities in this range. However, if indeed the contribution from the host emission is present in MIRI (see Section \ref{subsec:alt_models}), these measurements would be lower limits for CEERS 10444 and 13318.  Further, even if this model is applicable to most of the LRD population, these will likely have a range of bolometric luminosities.  \par 

Nevertheless, assuming the above AGN models are applicable to the LRDs, we can use the bolometric luminosities to make estimates of the SMBH mass limits, growth rates, and e-folding times. For this calculation, we assume that the bolometric luminosity stems from SMBH growth, and is related to the SMBH mass limit by,
\begin{equation}
    M_\mathrm{BH} = 3.05\times 10^7~M_\odot\ \left( \frac{L_\mathrm{bol,AGN}}{10^{12}~L_\odot}\right) \lambda^{-1}.
\end{equation}
Where $\lambda$ is the Eddington Ratio and $L_\mathrm{bol,AGN}$ is the bolometric luminosity of the AGN. For the \prospector\ and \cigale\ AGN models we assume this to be $\sim90\%$ of the total bolometric luminosity based on the resulting AGN fractions from the modeling. This yields limits on the SMBH mass of $\sim 5 \times 10^6 - 5 \times 10^7 $~$M_\odot$ for $\lambda \sim 1$. Accordingly, the mass accretion rate is defined as, $\dot M_\mathrm{acc}$, by $L_\mathrm{bol,AGN} = \epsilon\ \dot M_\mathrm{acc} c^2$ \citep{Tucci2017}. $\epsilon$ is the radiative efficiency, which can span from $\epsilon \simeq 0.05 - 0.4$ on theoretical grounds \citep[and references therein]{Tucci2017} but has an average of $\simeq 0.1$ in AGN accretion models \citep{Davis2011,Wu2013}. The SMBH growth rate is then $\dot{M} = (1-\epsilon)\dot M_\mathrm{acc}$, where inverting the mass--accretion rate yields 
\begin{equation}
    \dot M = 0.068~M_\odot~\mathrm{yr}^{-1}\ \left(\frac{L_\mathrm{bol,AGN}}{10^{12}~L_\odot}\right) \left( \frac{1-\epsilon}{\epsilon}\right).
\end{equation}
For our bolometric luminosities, this returns a limit of $\dot M \simeq 0.1$~$M_\odot$~yr$^{-1}$ for the dense gas and \cigale\ models, and $\dot M \simeq 1$~$M_\odot$~yr$^{-1}$  for \prospector\ assuming this average value of $\epsilon \simeq 0.1$. Lastly, the ratio of the SMBH mass to the growth rate yields a characteristic SMBH growth e-folding time, $\tau = M / \dot M$, which is approximately $\tau \sim 5\times 10^7 - 5\times 10^{8}$~yr for the values above. \par 

Taken together, these results indicate that understanding the growth of LRDs depend on three essential factors. First, it is imperative to understand if the models are {fully} representative of the emission mechanisms. Second, constraining the radiative efficiency and  Eddington ratio of LRDs is paramount to determining the growth timescales. Lastly, understanding the duration of the ``LRD-phase'' is fundamental for contextualizing the growth of these galaxies. Assuming this phase is $\sim 10^9$~yr, the LRDs could undergo 4 to 50 e-folding timescales, driving considerable SMBH growth.   \par 

\section{Summary and Conclusions} 
\label{sec:summary_concl}
In this paper, we present a comprehensive analysis on a sample of eight spectroscopically confirmed LRDs that have a combination of \jwst/NIRCam imaging, spectroscopic redshifts, and \jwst/MIRI imaging.  Six of the eight LRDs have \jwst/NIRSpec prism spectroscopy while the other two have redshifts from NIRCam/grism observations.  All have limits in the mid- and far-IR from archival \spitzer\ and \herschel\ imaging.   We jointly fit the multi--wavelength NIRCam and MIRI photometry, upper limits from \spitzer/MIPS and \herschel/PACS, and {NIRSpec}/prism data (when available)  with \prospector\ and \cigale. We model the LRDs with three cases (1) star-forming only models, (2) AGN-dominated models, where the AGN accounts for at minimum 50\% of the emission at 5500\AA; and (3) a composite model which allows the AGN contribution and star forming parameters to be free. We compute the BIC for each LRD for our modeling scenarios to assess which interpretation is best supported by the models.  Our primary findings are as follows. 

\begin{itemize}[itemsep=5pt]
\item We find that despite the fact that the LRDs are red in NIRCam F150W -- F444W $>$ 1, they are relatively blue NIRCam -- MIRI colors of $-0.5 < \mathrm{F444W - F770W} < 0.5$. However, the redder MIRI bands have colors spanning  $-1.0 < \mathrm{F1000W - F1500W} < 1.1$, which suggests the source of the observed mir-IR emission is not uniform within the population.  

\item Based on the BIC analysis from the SED model fits, we find the LRDs modeled with NIRSpec prism data tend to favor the AGN models.  This is true for both the \prospector\ and \cigale\ codes, where all the galaxies with prism data have at least some evidence of an AGN. This also underscores the importance for incorporating spectroscopy with sufficient signal-to-noise in the modeling to robustly evaluate the AGN contribution to LRDs.

\item Based on our SED modeling results from \cigale\ and \prospector, the MIRI data are not well reproduced by the torus as currently included in the \texttt{SKIRTOR} models nor the simplified \texttt{CLUMPY} models. In \prospector, the code is satisfied modeling the LRD emission using only the AGN continuum and star forming component to model the MIRI data. While \texttt{CIGALE} includes the torus component, we observe significant overestimation of the flux in the MIRI bands when this model is incorporated. These results imply that there is no torus in these objects with the physical assumptions in this model. However, it is equally plausible that standard torus models are insufficient for LRDs and may need revision.

\item Two of the LRDs in our sample, CEERS 10444 and 13318, show strong Balmer breaks, and Balmer absorption superimposed on their broad emission lines.  None of these features are fully reproduced with the models used by \cigale\ and \prospector.  For these galaxies, we consider alternative ``dense--gas'' models, that allow for very dense gas, heated to high temperatures by an AGN.  Using \texttt{CLOUDY} calculations for these models, we fit these to the imaging and NIRSpec data.  These models are able to reproduce shape of the data at wavelengths longer than the rest-frame Balmer break with dense gas ($\log n_H/\mathrm{cm^{-3}} = 10.5-11$), with temperature, $10^5-10^5.6$~K, high column densities, $\log N\mathrm{_H/cm^{-2}} = 24-25$, and moderate dust attenuation ($A(V) \simeq 0.5$~mag). These models also need an additional thermal source of 600--1300~K that accounts for 10--30\% of the total emission absorbed by the dust to account for the MIRI data.  However, these dense-gas models alone are insufficient to fully explain the emissions of these objects, as they fail to reproduce the observed narrow \oiii\ emission lines nor the rest-frame UV continuum emission. 

\item The models allow us to constrain the bolometric luminosities of the SED.   The averaged star-forming and AGN models from \prospector\ predict bolometric luminosities in excess of $\log L_\mathrm{bol}/L_\odot > 12$, but these would violate the limits in the far-IR and radio reported in the literature.  In these models the long wavelength emission stems from dust heated by star-formation.  This seems inconsistent with the data.  For the \cigale\ AGN models, the predicted  bolometric luminosities are $\log L_\mathrm{bol}/L_\odot = 11.3 \pm 0.4$.  However, most of these would also predict sub-mm flux densities in excess of the limits.  Only one of our objects has predicted sub-mm flux densities consistent with these limits.   The dense gas models predict bolometric luminosities of $\log L_\mathrm{bol}/L_\odot = 11.6-11.7$, assuming most of this emission occurs in dust heated to at least $T = 140$~K.  However to test these predictions requires measurements at far-IR and sub-mm wavelengths fainter than current observations.   
\end{itemize}

\begin{acknowledgments}
This work benefited from support from NASA/ESA/CSA
James Webb Space Telescope through the Space Telescope
Science Institute, which is operated by the Association of
Universities for Research in Astronomy, Incorporated, under NASA contract NAS5-03127. We acknowledge support from grant JWST-GO-03794.001. This work benefited from support from the George P. and Cynthia Woods Mitchell Institute for Fundamental Physics and Astronomy at Texas A\&M University. All of the data used in this article were obtained from the Mikulski Archive for Space Telescopes (MAST) at the Space Telescope Science Institute. FC acknowledges support from a UKRI Frontier Research Guarantee Grant (PI Cullen; grant reference EP/X021025/1). The publicly released mosaics from the CEERS Survey are available at ceers.github.
io/releases.html and on MAST via doi:10.17909/z7p0-8481.
\end{acknowledgments}

\software{astropy \citep{2013A&A...558A..33A,astropy_2018,Astropy_2022}, \cigale 
\citep{Boquien_2019}, \prospector\ \citep{Leja_2019,Johnson2021}, photutils \citep{Bradley2016}  Matplotlib \citep{Hunter2007}, pandas \citep{jeff_reback_2022_6408044}, \textsc{WebbPSF} \citep{Perrin2012}, JWST Calibration Pipeline \citep{ Bushouse2022, Bushouse2024, 2024Bushouse13.4} }
\clearpage
\begin{appendix} \label{sec:appendix}

\section{Line Fitting and Balmer Break Measurements}\label{sec:linefit}
Here we present the results from our fit of the observed lines in the \textit{NIRSpec}/G395M data with the exception of 3153, where the line fits use the \textit{NIRSpec}/prism data. We model the continuum as a linear component, the narrow lines with a single Gaussian, and broad lines as a two component Gaussian (one with a narrow component and one with a broad component, see Section~\ref{sec:methods}). We fit the widths of the Balmer emission lines separately from that of the metal lines as we expect these to differ \citep{wang2024}. We use LMFIT \citep{Newville2024} to model the emission lines, which performs non-linear optimization using the optimization methods of \textsc{scipy.optimize} \citep{Virtanen2020} using the Levenberg-Marquardt method.  \par

\begin{deluxetable}{ccccc}[h]
\tablecaption{Emission Line Fits \label{tab:linefits}}
\tablewidth{0pt}
\tablehead{
\colhead{CEERS ID} & \colhead{Lines Present} & \colhead{FWHM$_{broad}$ [km~s$^{-1}$]}& \colhead{FWHM$_{narrow}$ [km~s$^{-1}$]} & \colhead{Flux[$10^{-17}$erg$s^{-1}cm^{-2}$]} 
}
\startdata
3153 & \ha  & 3548{$\pm$}217 & 654{$\pm$}55 &4.49{$\pm$}0.55  \\
 &  \hb & \nodata & 654{$\pm$}55 & 0.09{$\pm$}0.02\\ [4pt]
10444 & \ha & 3160{$\pm$}84& 817{$\pm$}8 & 85.37{$\pm$}18.88  \\
 &  \hb & \nodata & 817{$\pm$}8 & 2.98{$\pm$}0.27\\
 &  [\ion{O}{3}] ($\lambda$4959) & \nodata & 315{$\pm$}10& 2.36{$\pm$}0.08  \\
  &  [\ion{O}{3}] ($\lambda$5007) & \nodata & 315{$\pm$}10 & 7.02{$\pm$}0.25 \\
13135 &  \ha & 2450{$\pm$}114 & 633{$\pm$}56 & 12.23{$\pm$}0.98\\
 &  [\ion{O}{3}] ($\lambda$4959) & \nodata & 298{$\pm$}26 & 0.45{$\pm$}0.04  \\
  &  [\ion{O}{3}] ($\lambda$5007) & \nodata & 298{$\pm$}26 & 1.35{$\pm$}0.12 \\[4pt]
13318 & \ha & 3386{$\pm$}78 & 584{$\pm$}27 & 93.93{$\pm$}15.54 \\
 &  \hb & \nodata & 584{$\pm$}27 & 2.70{$\pm$}0.25  \\
  &  [\ion{O}{3}] ($\lambda$4959)& \nodata &  427{$\pm$}25 & 1.14{$\pm$}0.07 \\ 
   &  [\ion{O}{3}]($\lambda$5007) & \nodata  & 427{$\pm$}25 & 3.38{$\pm$}0.22 \\ [4pt]
2520 &  \hb & \nodata & 335{$\pm$}56 & 1.25{$\pm$}0.19\\
 &  [\ion{O}{3}] ($\lambda$4959) & \nodata & 249{$\pm$}8 & 2.53{$\pm$}0.08  \\
  &  [\ion{O}{3}] ($\lambda$5007) & \nodata & 249{$\pm$}8 & 7.53{$\pm$}0.26  \\[4pt]
20496 &  \ha & 1714{$\pm$}486 & 199{$\pm$}17 & 3.88{$\pm$}0.71 \\
 &  \hb & \nodata &  199{$\pm$}17 & 0.42{$\pm$} 0.07 \\
  &  [\ion{O}{3}] ($\lambda$4959)& \nodata & 284{$\pm$}9 & 1.18{$\pm$}0.04 \\
  & [\ion{O}{3}]($\lambda$5007) & \nodata & 284{$\pm$}9 & 3.51{$\pm$}0.12 \\
\hline \hline
\enddata
\end{deluxetable}

Additionally we measure the strength of the 
Balmer break for each object in our 
sample with the \textit{NIRSpec}/prism data 
using the definition \cite{Wang2024_} presented in Table~\ref{tab:breakmeas} below. 
For this work we define the strength of the Balmer 
Break as the average flux between (3620\AA, 3700\AA) and (4000\AA, 4100\AA), which differs from \cite{Balogh1999} to mitigate contamination from strong lines. We calculate this using 
\begin{equation}
    BB = \frac{\lambda_2^{blue}- {\lambda_1^{blue}}}{\lambda_2^{red}-{\lambda_1^{red}}} \frac{\int_{\lambda_1^{red}}^{\lambda_2^{red}} F_\nu ^{red}d\lambda}{\int_{\lambda_1^{blue}}^{\lambda_2^{blue}} F_\nu ^{blue}d\lambda},
\end{equation}
where $\lambda_2^{blue}, \lambda_1^{blue}, \lambda_2^{red}$, and $\lambda_1^{red}$ are 3620\AA, 3700\AA, 4000\AA, and 4100\AA,  and $F_\nu ^{blue}$ and $F_\nu ^{red}$ are the fluxes within their respective wavelength ranges. 

\begin{deluxetable}{cc}[h]
\tablecaption{Balmer Break Measurements \label{tab:breakmeas}}
\tablewidth{0pt}
\tablehead{
\colhead{CEERS ID~~~~~} & \colhead{~~~~~Balmer Break Strength}
}
\startdata
3153 &  1.02{$\pm$}0.25\\[0.13cm]
10444 &  2.88{$\pm$}0.23 \\[0.13cm]
13135 &  1.68{$\pm$}0.54 \\[0.13cm]
13318 &  1.68{$\pm$}0.16 \\[0.13cm]
2520 &  0.91{$\pm$}0.14 \\[0.13cm]
20496 &  1.16{$\pm$}0.26 \\[0.13cm]
\hline \hline
\enddata
\end{deluxetable}

\section{SED Fitting Parameters and Results}\label{sec:sedparams}
\begin{figure*}
    \centering
    \includegraphics[scale=0.35]{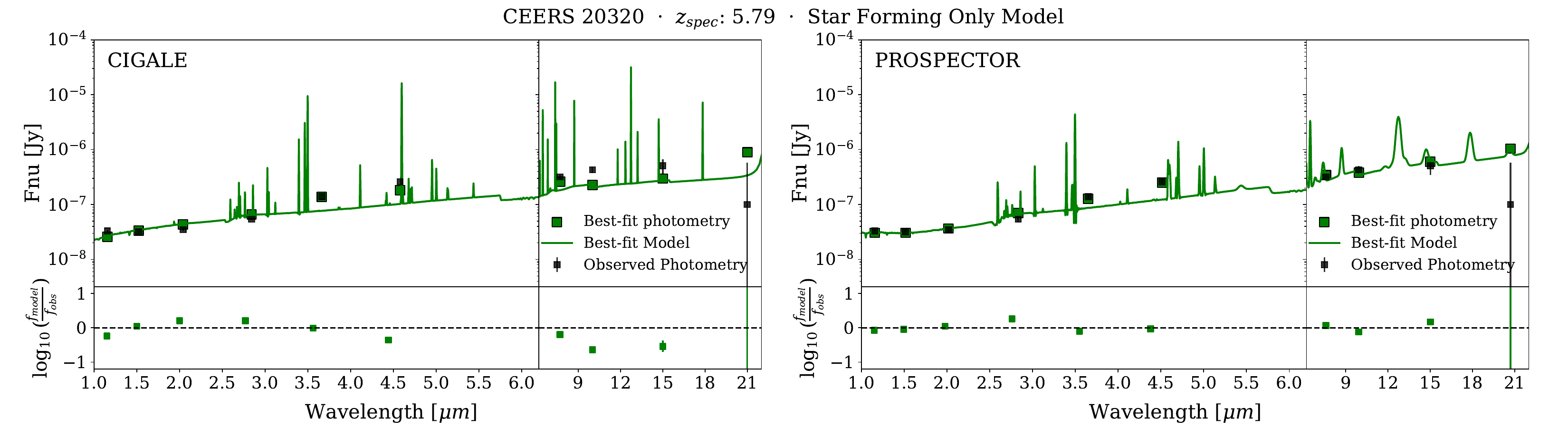}
    \includegraphics[scale=0.35]{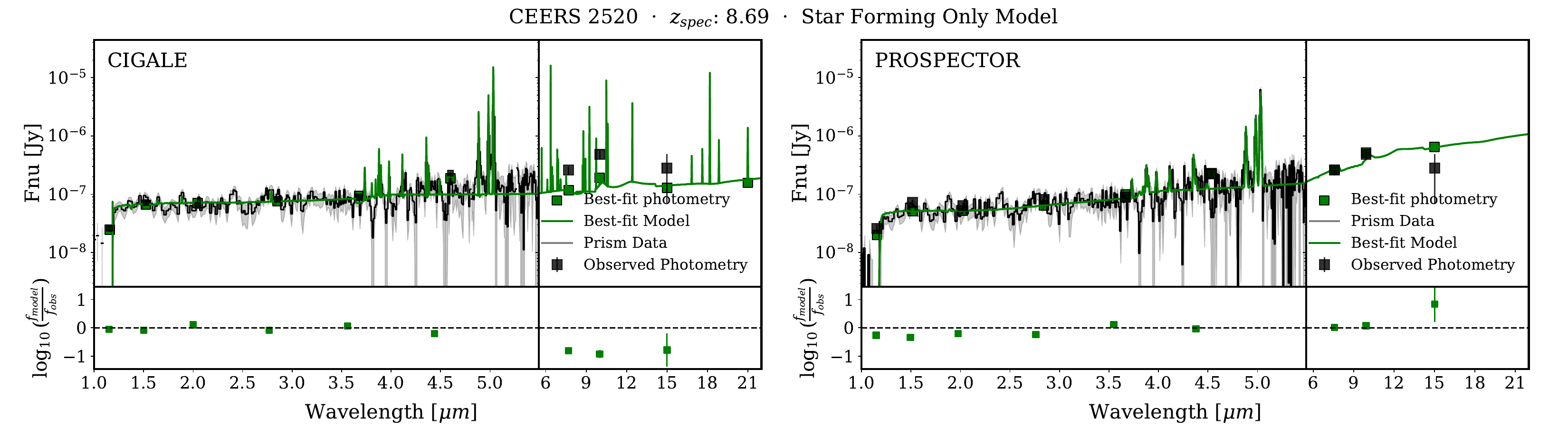}
    \includegraphics[scale=0.35]{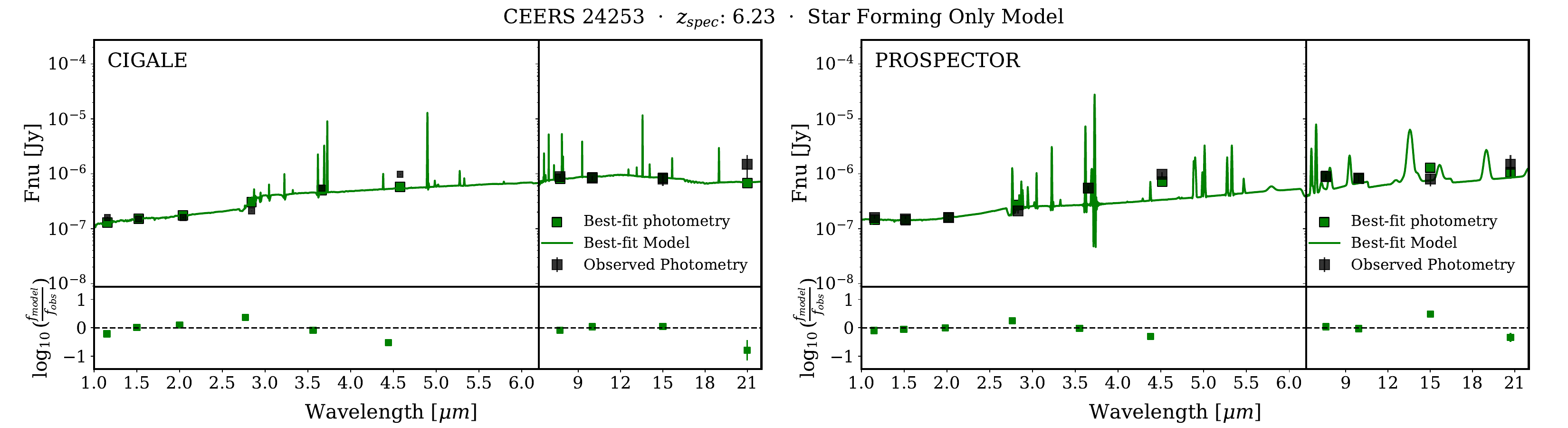}
    \includegraphics[scale=0.35]{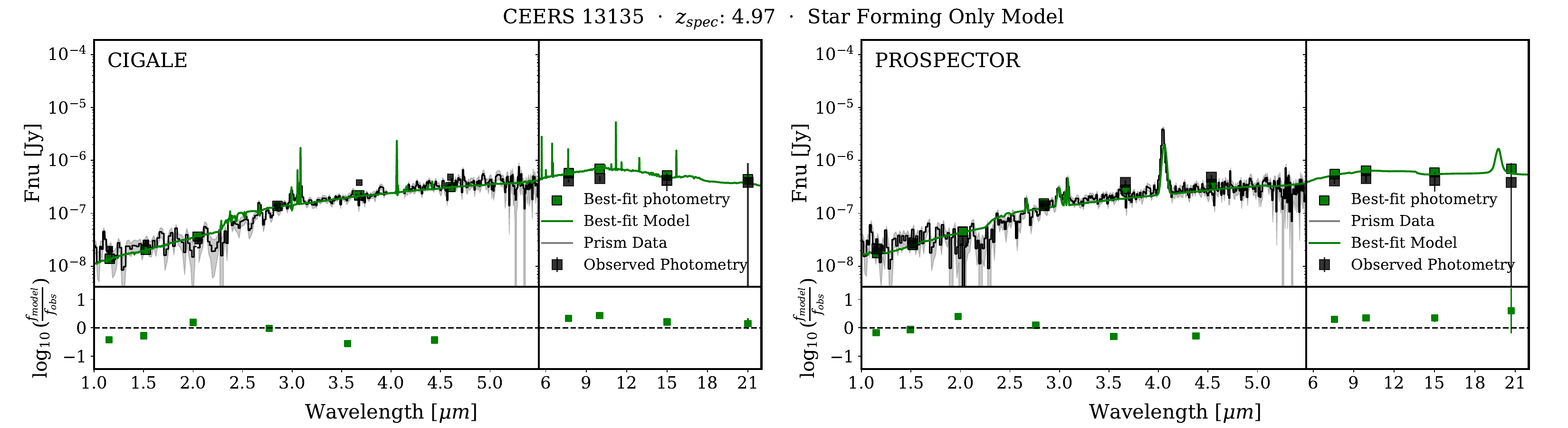} 
    \caption{Like Figure~\ref{fig:sf_only}, but showing the fitting results for the star-forming only models for the remaining four LRDs in our sample.  \label{fig:sf_only_appendix}}
\end{figure*}

\begin{figure*}
    \centering
    \includegraphics[scale=0.35]{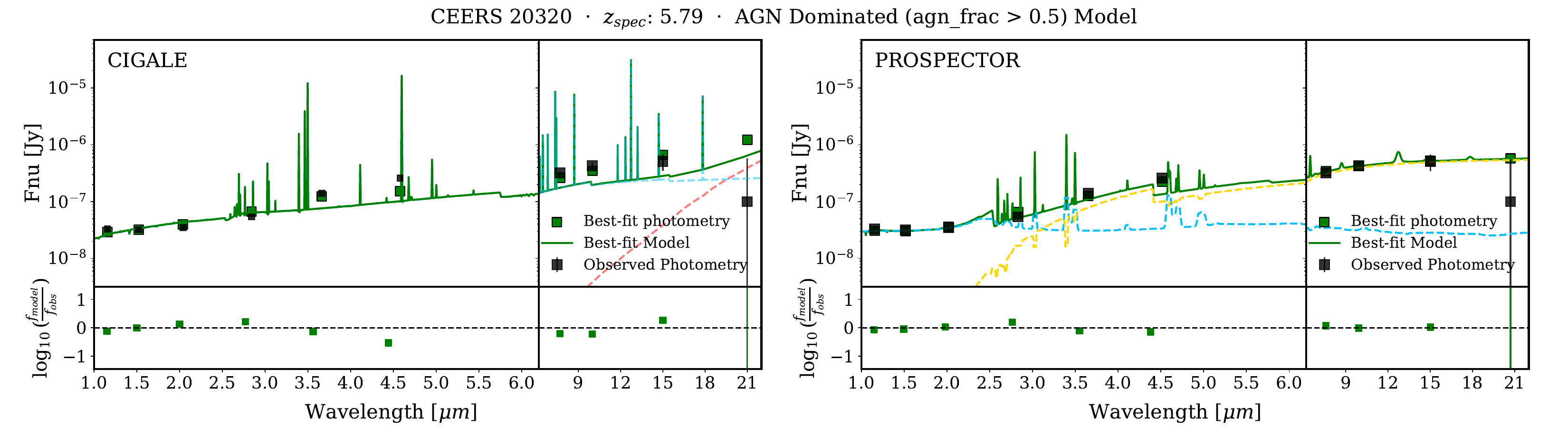}
    \includegraphics[scale=0.35]{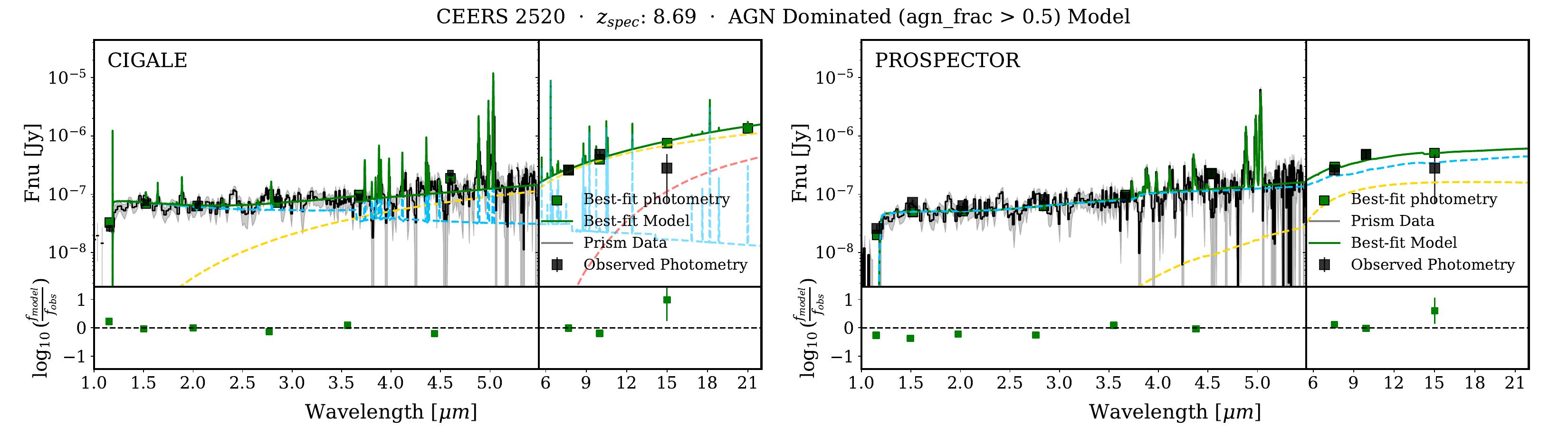}
    \includegraphics[scale=0.35]{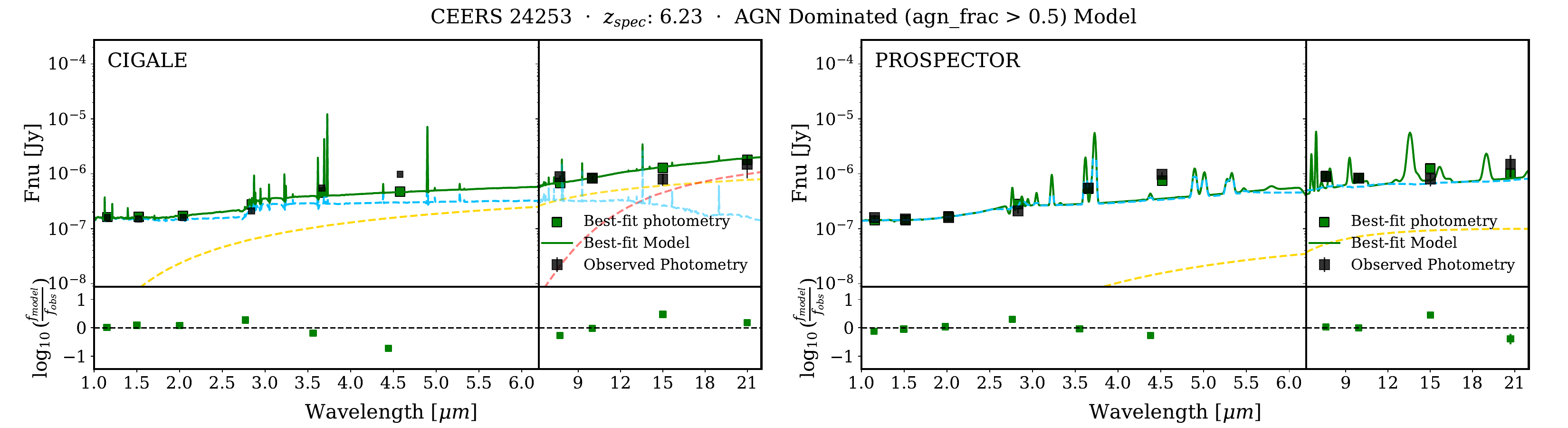}
    \includegraphics[scale=0.35]{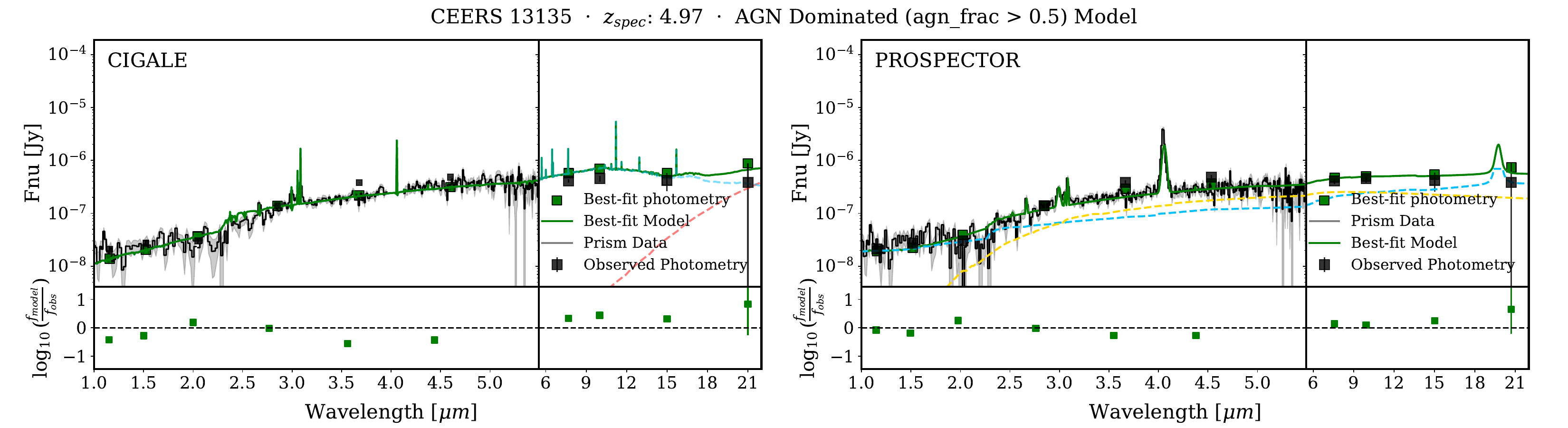}
    \caption{Like Figure~\ref{fig:forceagn}, but showing the fitting results for the AGN-dominated models for the remaining four LRDs in our sample. \label{fig:agn_appendix}}
\end{figure*}

\begin{figure*}
    \centering
    \includegraphics[scale=0.35]{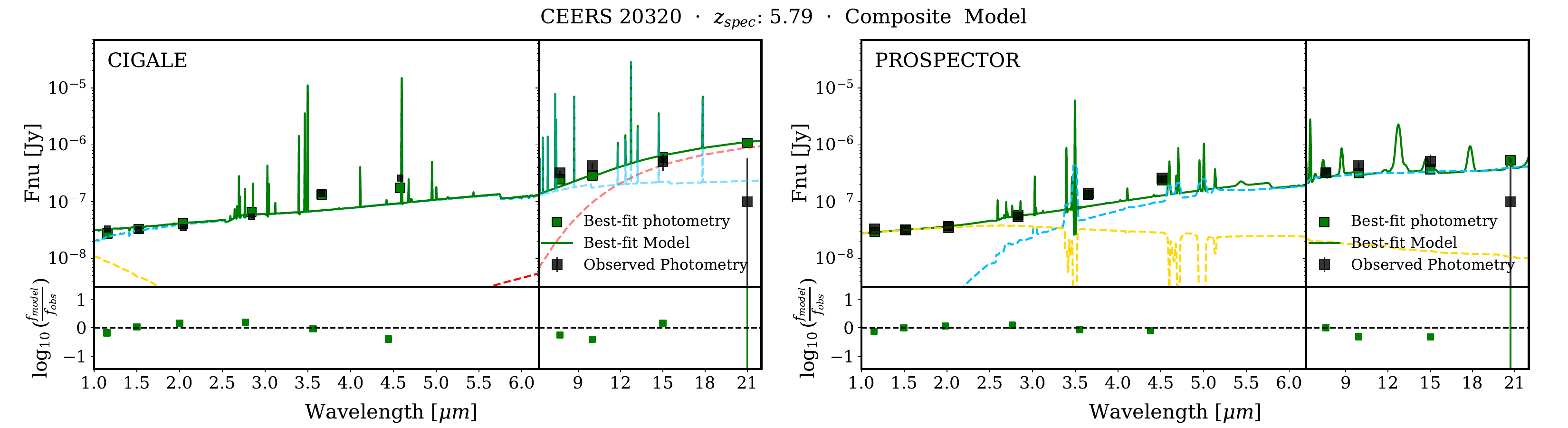}
    \includegraphics[scale=0.35]{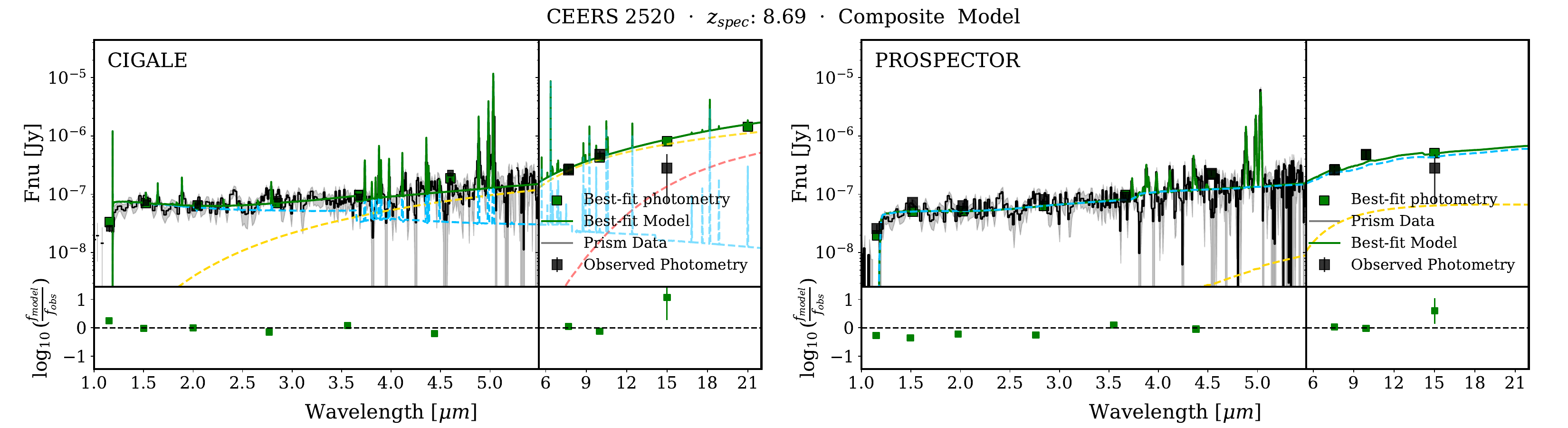}
    \includegraphics[scale=0.35]{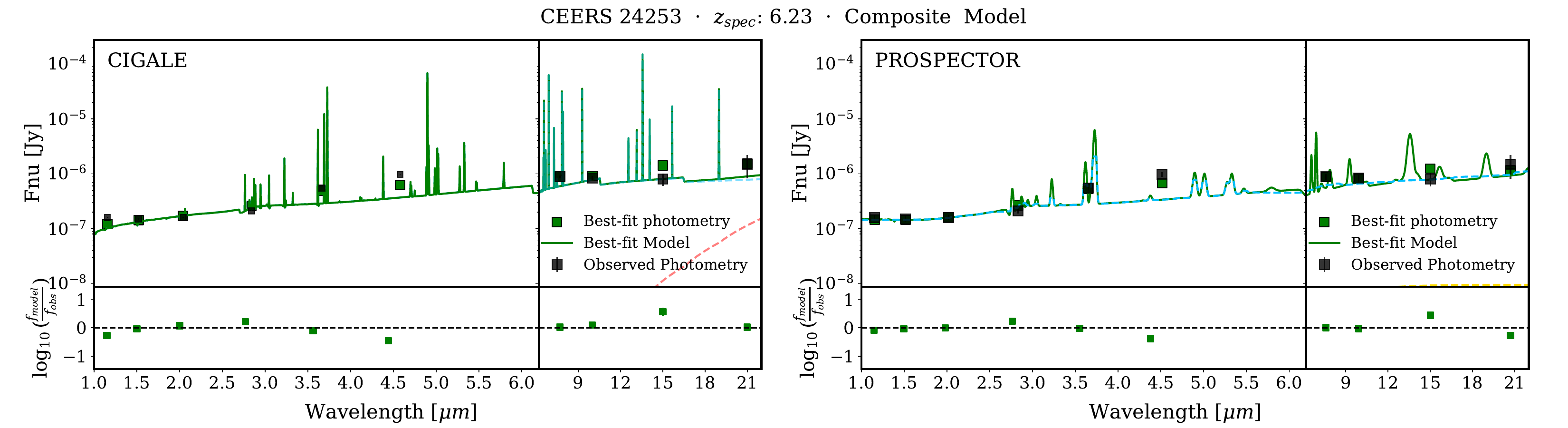}
    \includegraphics[scale=0.35]{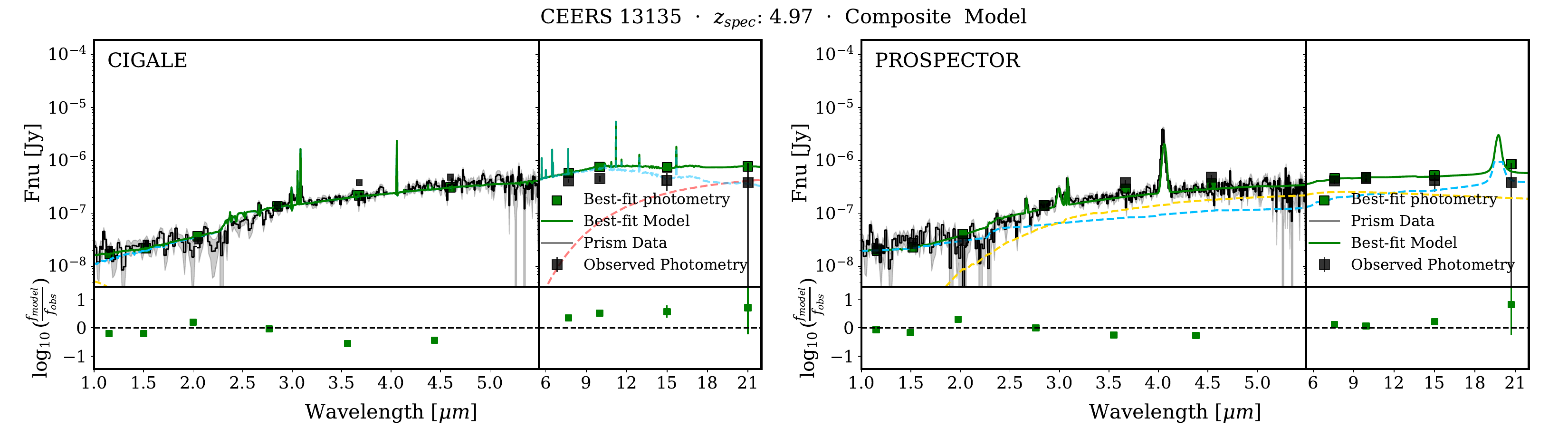}
    \caption{Like Figure~\ref{fig:composite}, but showing the fitting results for the composite AGN and star-forming models for the remaining four LRDs in our sample.  \label{fig:comp_appendix}}
\end{figure*}

In this Section we display the parameters to model each component in \cigale\ and \prospector\ shown in Table~\ref{tab:cigale_param} and Table~\ref{tab:prospy_param}, respectively. In addition we include the SED modeling results for the rest of our sample for the star forming only models in Figure~\ref{fig:sf_only_appendix}, the AGN dominated model in Figure~\ref{fig:agn_appendix}, and the composite star-forming and AGN models in Figure~\ref{fig:comp_appendix}.
\begin{deluxetable}{lll}[ht]
\tabletypesize{\footnotesize}
\tablecaption{\textsc{CIGALE} parameters used for SED fitting \label{tab:cigale_param}}
\tablewidth{0pt}
\tablehead{
\colhead{\raggedleft{Module}} & \colhead{Parameter} & \colhead{Input Value(s)}
}
\startdata
Periodic SFH: & \texttt{age} [Myr] & {10, 100, 200, 400, 600, 800, 1000}\\
SFR(t) \(\propto \frac{t}{\tau ^2} \times exp (\frac{t}{- \tau})\) & \texttt{delta\_bursts} [Myr] & {10, 100} \\
\hline
Simple stellar population: & IMF & \cite{Chabrier2003} \\  
\cite{Bruzual2003}& &  \\ \hline
Nebular & \texttt{metallicity} [$Z_\odot$] & {0.0004, 0.004,0.008, 0.02}  \\ 
& \texttt{logU} & {-4.0,-3.0, -2.0, -1.0} \\ \
& \texttt{zgas} & {0.001, 0.004,0.011, 0.025,0.03, 0.051}0 \\ 
& \texttt{ne} & {10, 100, 1000} \\ \hline
Dust attenuation: & \texttt{E\_BV\_lines} & {1e-4, 0.001, 0.010, 0.050,  0.10, 0.15, 0.20, 0.30, 0.50, 1.00} \\ 
\cite{Calzetti2000} & \texttt{Ext\_law\_emission\_lines} & SMC \\ \hline
Dust Emission: & \texttt{qpah} & {0.47, 1.12, 2.50,3.19, 4.58, 5.26, 6.63}\\
\cite{Draine2014} & \texttt{umin} & {0.1, 1, 5, 10, 15, 25}\\
{       }& \texttt{gamma} & {0.25, 0.54, 0.75} \\ 
{       }& \texttt{alpha}& {2.4} \\ \hline
AGN Emission: & \texttt{oa} & {40}\\
{       }& \texttt{t} & {3, 7, 11} \\ 
\cite{Stalevski2012, Stalevski2016} & \texttt{i} & 20, 40, 60, 80 \\
{       }& \texttt{disk\_type} & \cite{Schartmann2005} \\ 
{       }& \texttt{delta} & {-5.0,0,0.5} \\
{       }& \texttt{fracAGN} & {0.1, 0.25, 0.35, 0.5, 0.75, 0.85, 0.9} \\
{       }& \texttt{EBV} &{0,0.2,0.4,0.6}\\
\hline \hline
\enddata
\tablecomments{The default \cigale\ values were used for parameters not listed in this table.}
\end{deluxetable}

\begin{deluxetable}{lll}
\tabletypesize{\footnotesize}
\tablecaption{\prospector\ parameters used for SED fitting \label{tab:prospy_param}}
\tablewidth{0pt}
\tablehead{
\colhead{\raggedleft{Module}} & \colhead{Parameter} & \colhead{Prior}}
\startdata
Stellar/Nebular: & IMF & \cite{Chabrier2003} \\ 
{    }& \texttt{gas\_logu} & Uniform(min=-4.0, max=-1) \\ 
{    }& \texttt{gas\_logz} & Uniform(min=-2, max=0.5) \\ 
{    }& \texttt{logzsol} & Uniform(min=-1.98, max=0.19 \\ \hline
Dust attenuation: & \texttt{dust\_index} & Uniform(min=-1.0, max=0.4) \\ 
\cite{Kriek2013} & \texttt{dust2} & Normal(min=0.0, max=4.0, $\mu$=0.3, $\sigma$=1.0) \\ \hline
Dust Emission: & \texttt{log\_duste\_gamma}& Normal(min=-4.0, max=0.0, $\mu$=-2.0, $\sigma$=1.0 \\
\cite{Draine2007} & \texttt{duste\_qpah} & Normal(min=0.0, max=7.0, $\mu$=2.0, $\sigma$=2.0)\\
{} & \texttt{umin} & Normal(min=0.1, max=25.0, $\mu$=1.0, $\sigma$=10.0) \\ \hline
AGN Accretion Disk: & \texttt{agn\_frac} & Uniform(min=0.1, max=1)\\
\cite{Temple2021} & \texttt{agn\_dust4} & Uniform(a=0, b=4)\\
{} &\texttt{agn\_dust\_index} & Uniform(min= -1.8, max=-.8)\\ \hline
AGN Torus: & \texttt{log\_fagn} & Uniform(min=$\log$(0.1), max=$\log($3)) \\
\citep{Nenkova2008} & \texttt{log\_agn\_tau} & Uniform(min=$\log$(5.0), max=$\log$(150.0)) \\ \hline
\hline \hline
\enddata
\tablecomments{The default \prospector\ values were used for parameters not listed in this table. For more information on the listed FSPS parameters see \href{https://github.com/cconroy20/fsps/blob/master/doc/MANUAL.pdf}{this} link.}
\end{deluxetable}

\clearpage

\section{Sample Presentation Cont.}\label{sec:postage_stamps_cont}
In this Section of the Appendix,  we present the \textit{NIRSPec/}G395M and \textit{NIRCam}/grism spectra for all LRDs in our sample, except for 10444, which is shown in Figure~\ref{fig:sampleshow} above.

\begin{figure*}
\centering
\includegraphics[scale=0.35]{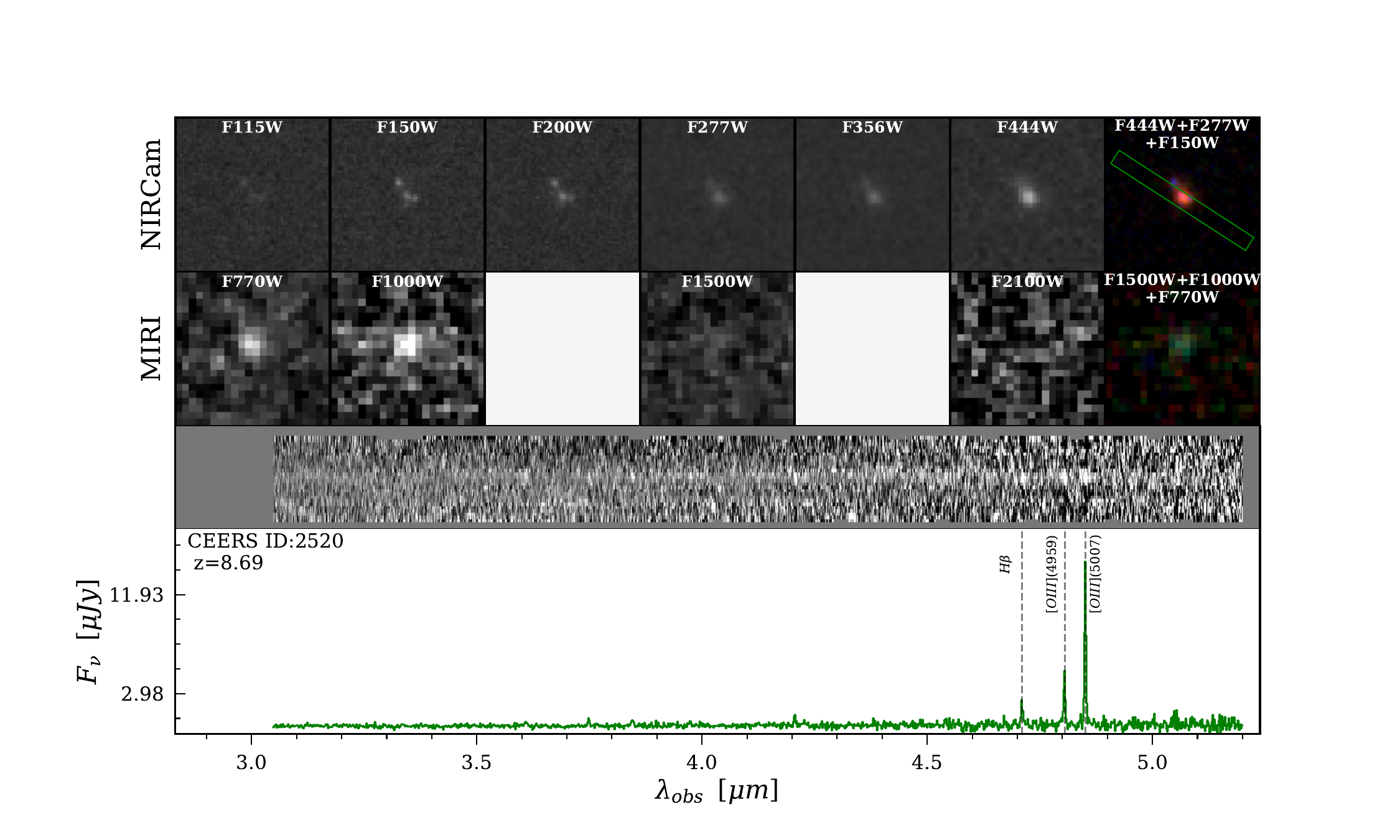}
\caption{\jwst/NIRCam (first row) and MIRI (second row) 2"x2" postage 
stamps. The last postage stamp in each row are the RGB images with 
F444W (R) + F277W (G) + F150W (B) for NIRCam, and F1500W (R) + F1000W 
(G) + F770W (B) for MIRI. In the NIRCam RGB image, we provide 
approximate NIRSPec slit positions with the green rectangle. The bottom 
two rows are the 2D and 1D \textit{NIRSpec}/G395M for 2520 and prism 3153 data for 3153, with the observed emission lines labeled. \label{fig:sampleshow_cont}}
\end{figure*}

\begin{figure*}
\centering
\includegraphics[scale=0.45]{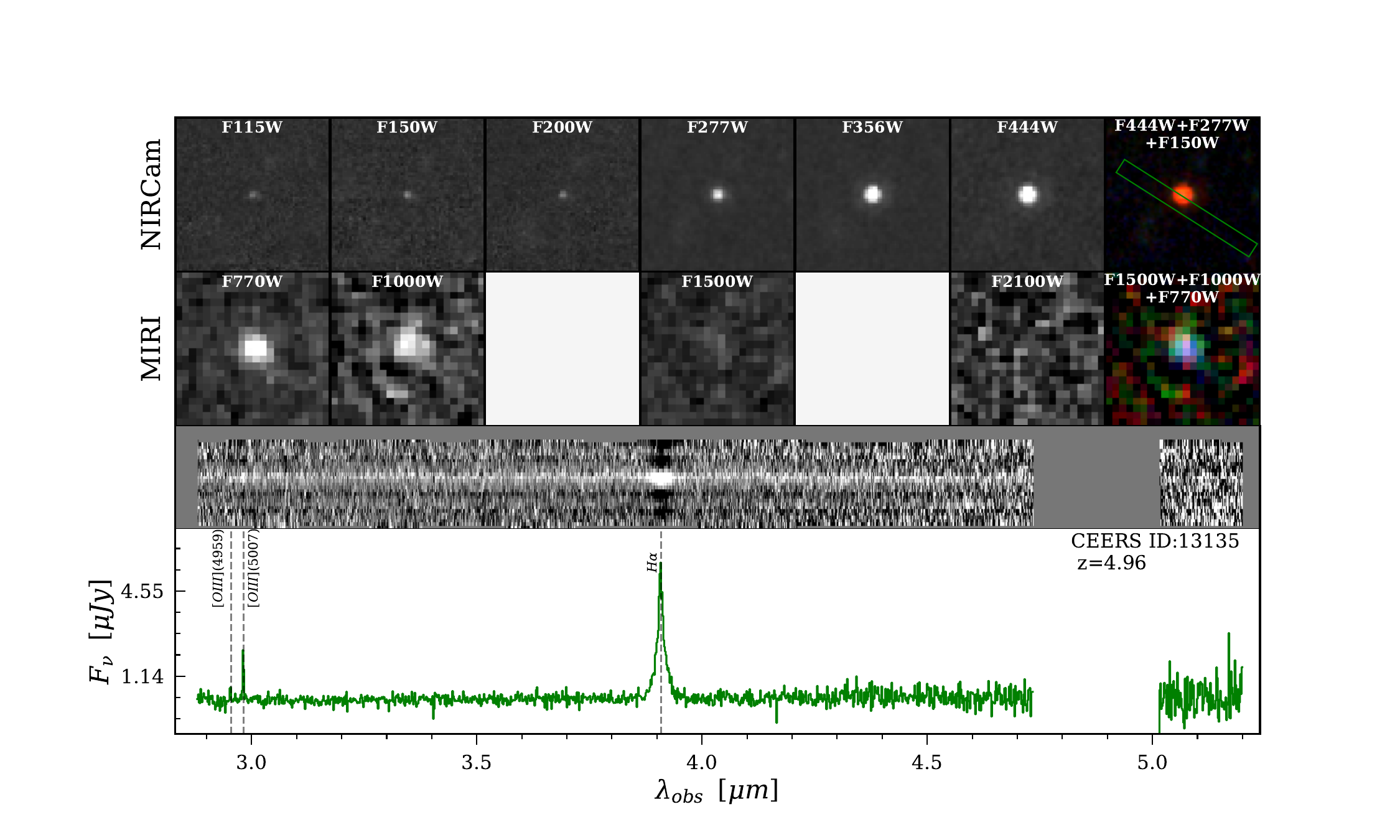}
\includegraphics[scale=0.45]{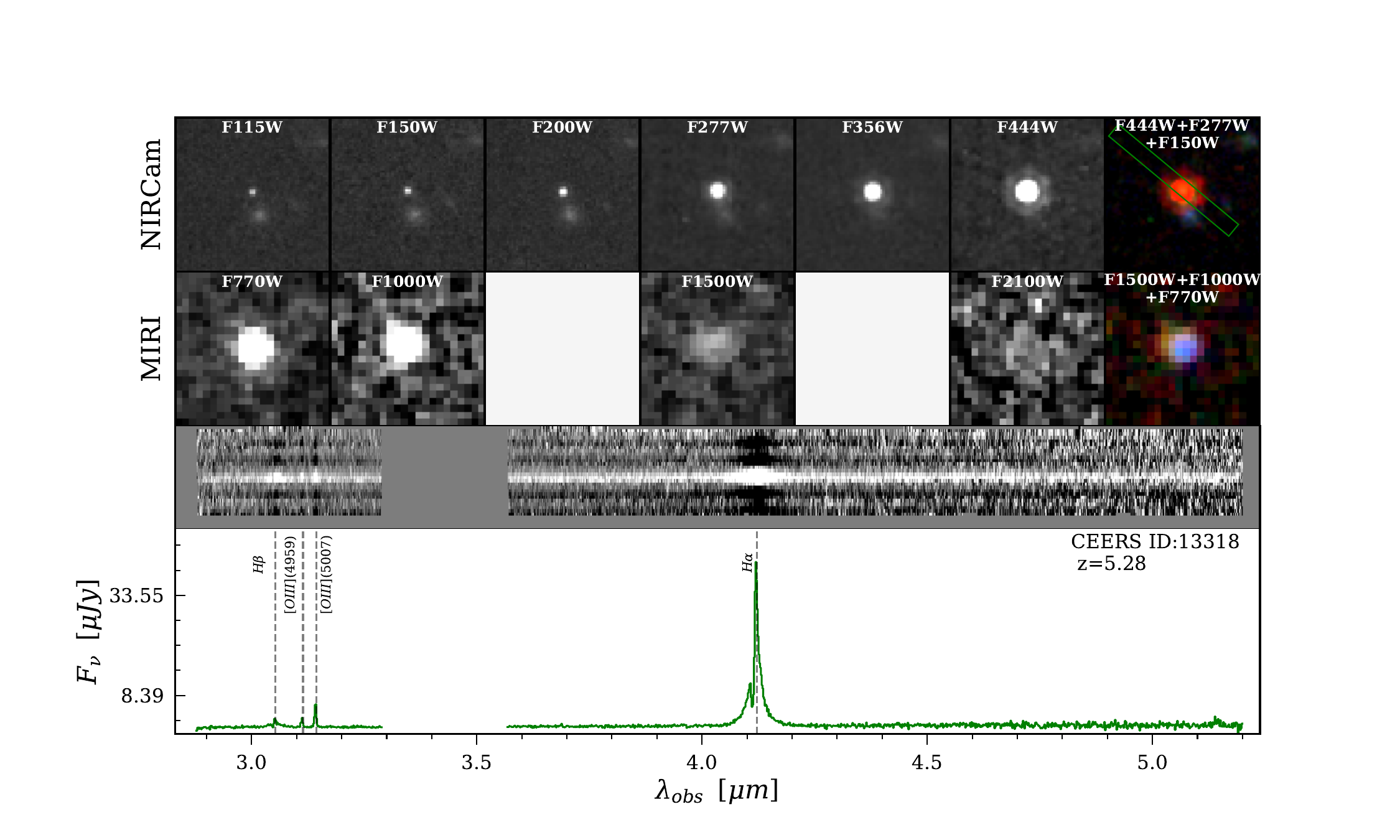}
\caption{\jwst/NIRCam (first row) and MIRI (second row) 2"x2" postage 
stamps. The last postage stamp in each row are the RGB images with 
F444W (R) + F277W (G) + F150W (B) for NIRCam, and F1500W (R) + F1000W 
(G) + F770W (B) for MIRI. In the NIRCam RGB image, we provide 
approximate NIRSPec slit positions with the green rectangle. The bottom 
two rows are the 2D and 1D \textit{NIRSpec}/G395M data for 13135 and 13318, with the observed emission lines labeled. \label{fig:sampleshow_cont}}
\end{figure*}

\begin{figure*}
\centering
\includegraphics[scale=0.45]{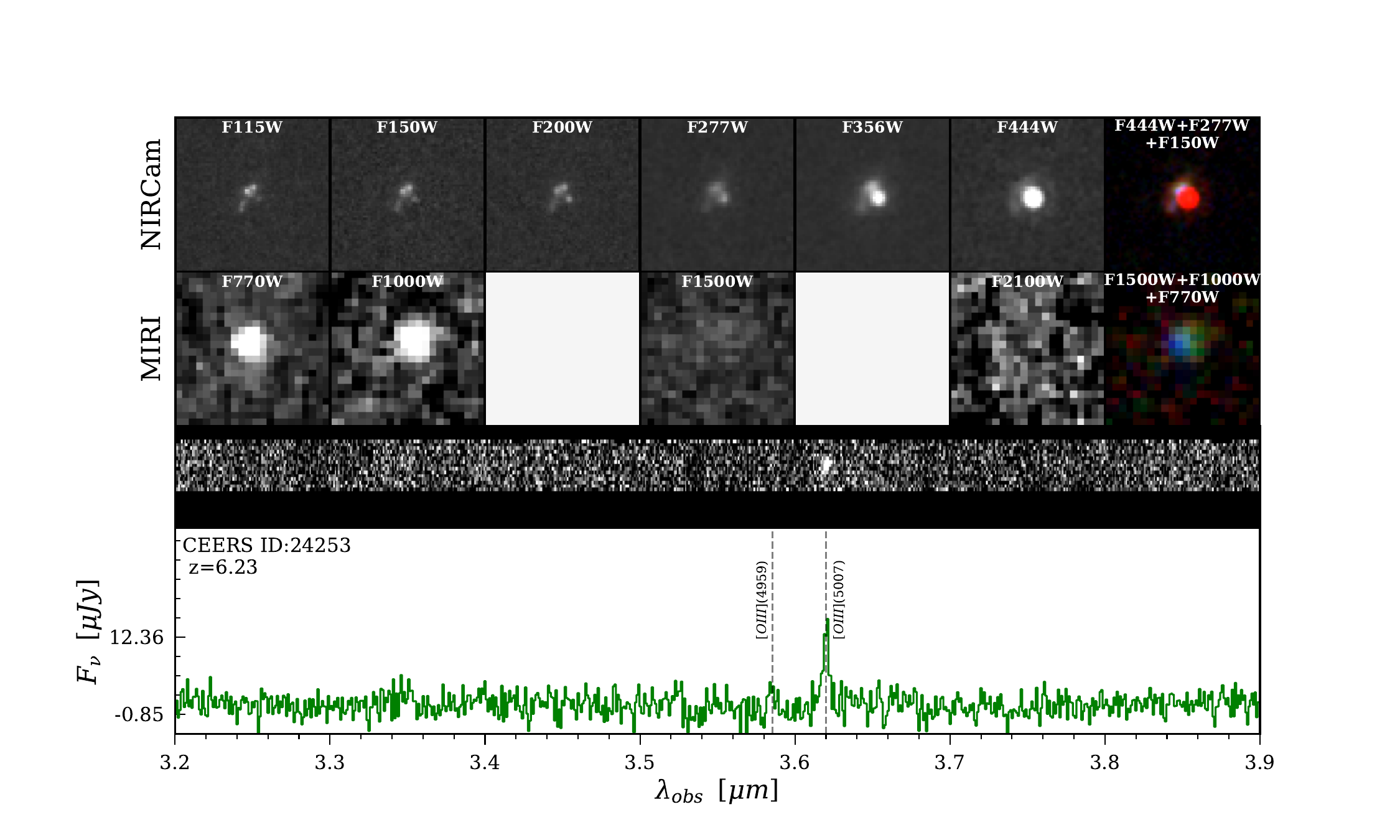}
\includegraphics[scale=0.45]{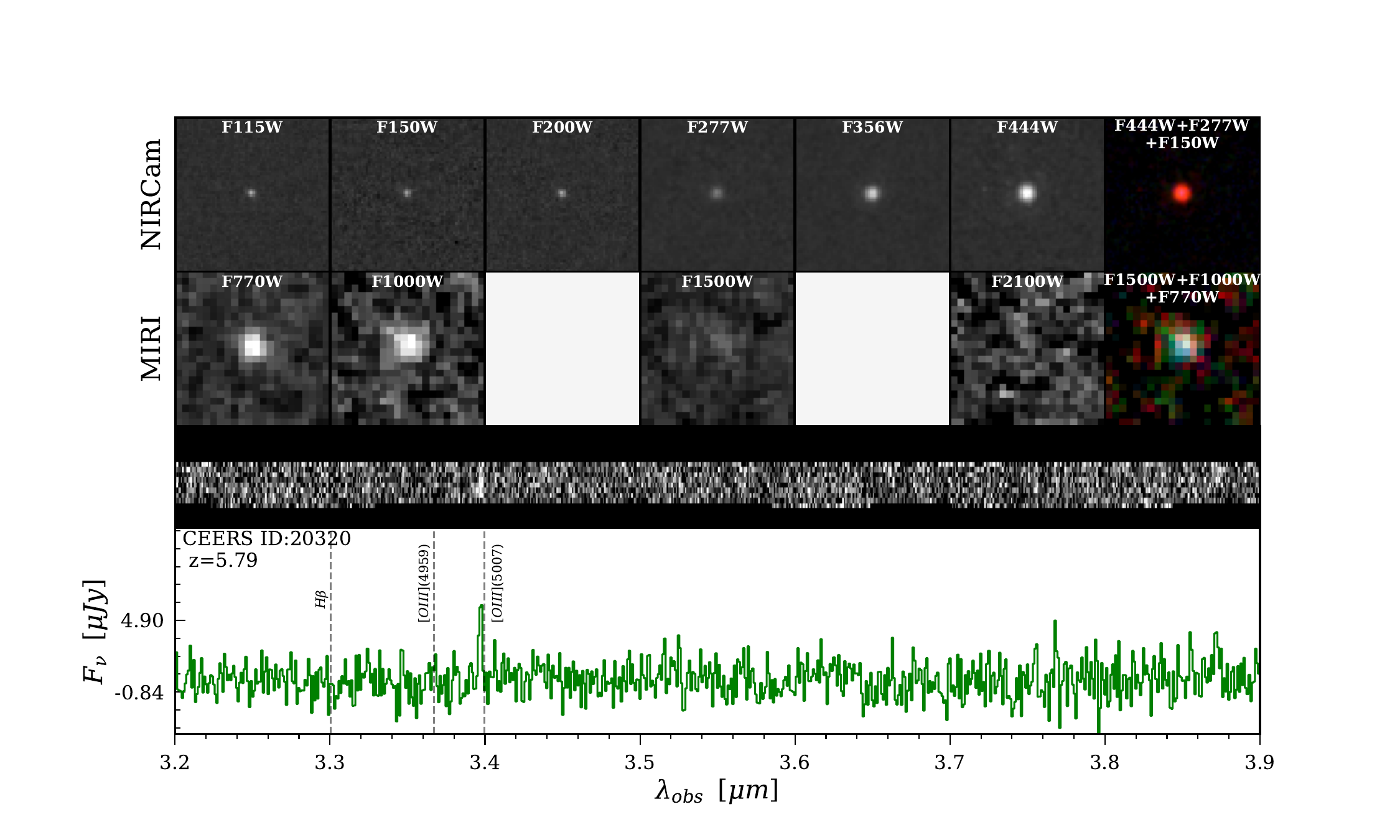}
\caption{\jwst/NIRCam (first row) and MIRI (second row) 2"x2" postage 
stamps. The last postage stamp in each row are the RGB images with 
F444W (R) + F277W (G) + F150W (B) for NIRCam, and F1500W (R) + F1000W 
(G) + F770W (B) for MIRI. The bottom two rows are the 2D and 1D \textit{NIRCam}/grism data for 24253 and 20320, with the observed emission lines labeled. \label{fig:sampleshow_cont}}
\end{figure*}

\begin{figure*}
\centering
\includegraphics[scale=0.45]{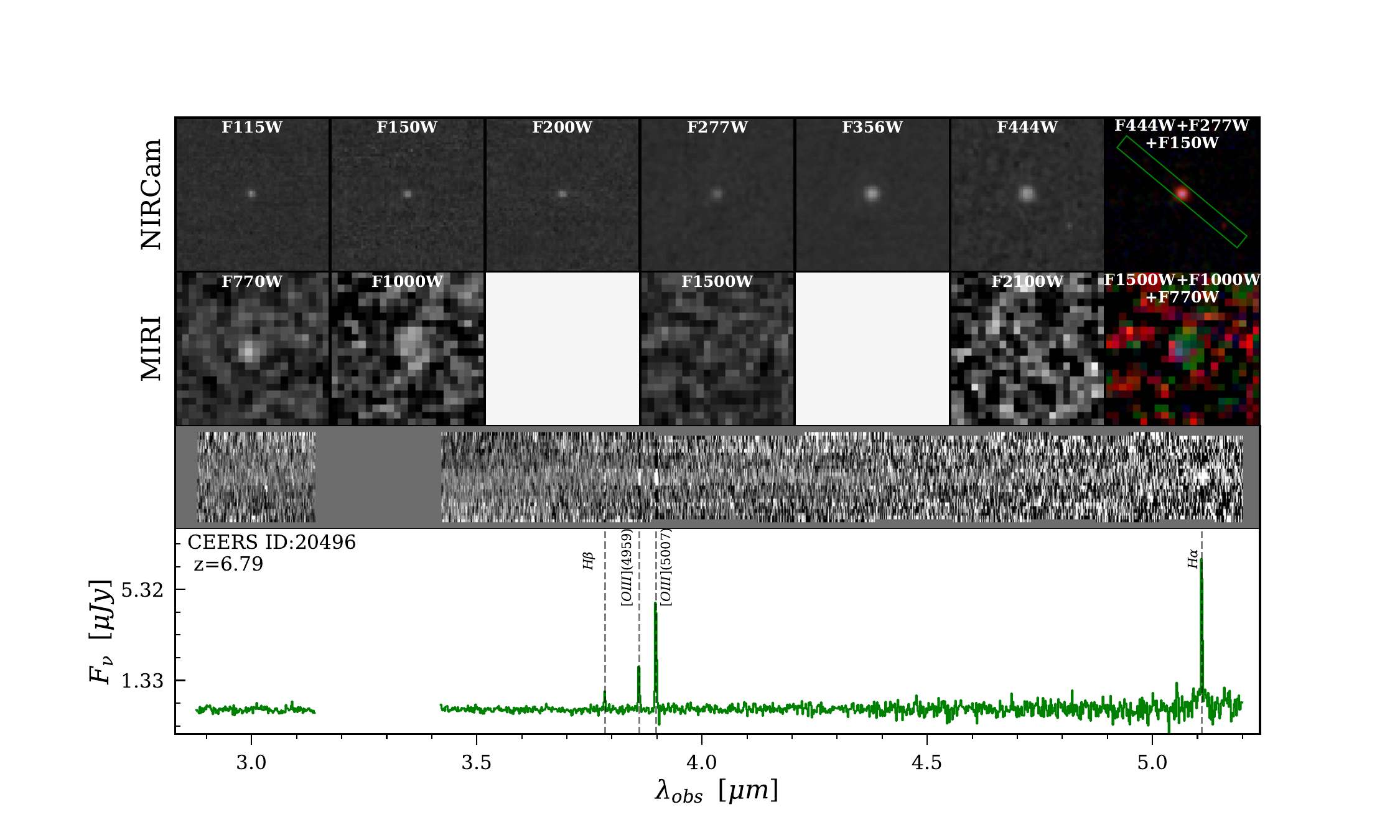}
\includegraphics[scale=0.45]{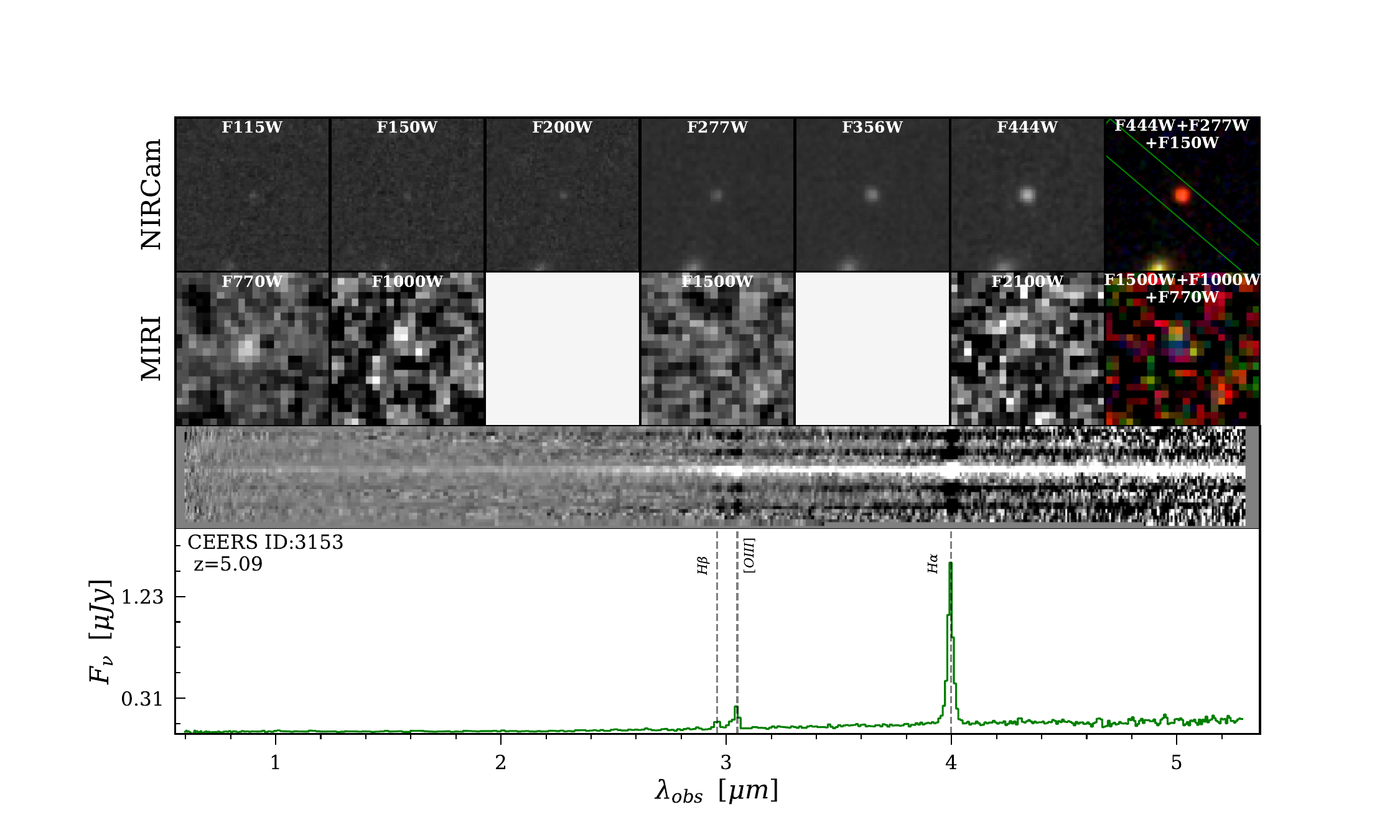}
\caption{\jwst/NIRCam (first row) and MIRI (second row) 2"x2" postage 
stamps. The last postage stamp in each row are the RGB images with 
F444W (R) + F277W (G) + F150W (B) for NIRCam, and F1500W (R) + F1000W 
(G) + F770W (B) for MIRI. In the NIRCam RGB image, we provide 
approximate NIRSPec slit positions with the green rectangle. The bottom 
two rows are the 2D and 1D \textit{NIRSpec}/G395M data for 20496, with the observed emission lines labeled. \label{fig:sampleshow_cont}}
\end{figure*}

\clearpage
\end{appendix}

\bibliography{citedpapers}{}
\bibliographystyle{aasjournalv7}
\end{CJK}
\end{document}